\documentstyle[preprint,prabib,aps]{revtex}

\newtheorem{corollary}{Corollary}
\newtheorem{theorem}{Theorem}
\newtheorem{lemma}{Lemma}

\begin{document}
\draft
\date{\today}
\title{A New Method of Constructing Black Hole \\
Solutions in Einstein and 5D Gravity }
\author{Sergiu I. Vacaru \thanks{
E-Mails : sergiu$_{-}$vacaru@yahoo.com,\ sergiuvacaru@venus.nipne.ro}}
\address{Physics Department, CSU Fresno,\ Fresno, CA 93740-8031, USA, \\
and \\
Centro Multidisciplinar de Astrofisica - CENTRA, Departamento de Fisica,\\
Instituto Superior Tenico, Av. Rovisco Pais 1, Lisboa, 1049-001, Portugal}
\maketitle

\begin{abstract}
It is formulated a new 'anholonomic frame' method of constructing
 exact solutions of Einstein equations with off--diagonal
metrics in 4D and 5D gravity. The previous approaches and results
\cite{v,v2,vf,v1} are summarized and generalized  as three
theorems which state the conditions when two types of ansatz
result in integrable gravitational field equations.  There are
constructed and analyzed different classes of anisotropic and/or
warped vacuum 5D and 4D metrics describing ellipsoidal black
holes with static anisotropic horizons and possible anisotropic
gravitational polarizations and/or running  constants. We
conclude that warped metrics can be defined in 5D vacuum gravity
without postulating any brane configurations with specific energy
momentum tensors. Finally, the 5D and 4D anisotropic Einstein
spaces with cosmological constant are investigated. \vskip5pt

Pacs 04.50.+h, 11.25.M, 11.10.Kk, 12.10.-g
\end{abstract}






\section{Introduction}

During the last three years large extra dimensions and brane worlds attract
a lot of attention as possible new paradigms for gravity, particle physics
and string/M--theory. As basic references there are considered Refs. \cite
{stringb}, for string gravity papers, the Refs. \cite{arkani}, for extra
dimension particle fields, and gravity phenomenology with effective Plank
scale and \cite{rs}, for the simplest and comprehencive models proposed by
Randall and Sundrum (in brief, RS; one could also find in the same line some
early works \cite{akama} as well to cite, for instance, \cite{shir} for
further developments with supersymmetry, black hole solutions and
cosmological scenaria).

The new ideas are based on the assumption that our Universe is realized as a
three dimensional (in brief, 3D) brane, modeling a 4D pseudo--Riemannian
spacetime, embedded in the 5D anti--de Sitter ($AdS_5$) bulk spacetime. It
was proved in the RS papers \cite{rs} that in such models the extra
dimensions coud be not compactified (being even infinite) if a nontrivial
warped geometric configuration is defined. Some warped factors are essential
for solving the mass hierarchy propoblem and localization of gravity which
at low energies can ''bound'' the matter fields on a 3D subspace. In
general, the gravity may propagate in extra dimensions.

In connection to modern string and brane gravity it is very
important to develop new methods of constructing  exact solutions
of gravitational field equations in the bulk of extra dimension
spacetime and to develop new applications in particle physics,
astrophysics and cosmology. This paper is devoted to elaboration
of a such method and investigation of new classes of anisotropic
black hole solutions.

In higher dimensional gravity much attention has been paid to the
off--diagonal metrics beginning the Salam, Strathee and Peracci works \cite
{sal} which showed that including off--diagonal components in higher
dimensional metrics is equivalent to including $U(1),SU(2)$ and $SU(3)$
gauge fields. They considered a parametrization of metrics of type
\begin{equation}
g_{\alpha \beta }=\left[
\begin{array}{ll}
g_{ij}+N_i^aN_j^bh_{ab} & N_j^eh_{ae} \\
N_i^eh_{be} & h_{ab}
\end{array}
\right]  \label{salam}
\end{equation}
where the Greek indices run values $1,2,...,n+m,$ the Latin indices $%
i,j,k,...$ from the middle of the alphabet run values $1,2,...,n$ (usually,
in Kaluza--Klein theories one put $n=4)$ and the Latin indices from the
beginning of the alphabet, $a,b,c,...,$ run values $n+1,n+2,...,n+m$ taken
for extra dimensions. The local coordinates on higher dimensional spacetime
are denoted $u^\alpha =\left( x^i,y^a\right) $ which defines respectively
the local coordinate frame (basis), co--frame (co--basis, or dual basis)
\begin{eqnarray}
\partial _\alpha &=&\frac \partial {\partial u^\alpha }=\left( \partial _i=%
\frac \partial {\partial x^i},\partial _a=\frac \partial {\partial y^a}%
\right) ,  \label{pder} \\
d^\alpha &=&du^\alpha =\left( d^i=dx^i,d^a=dy^a\right) .  \label{pdif}
\end{eqnarray}
The coefficients $g_{ij}=g_{ij}\left( u^\alpha \right) ,h_{ab}=h_{ab}\left(
u^\alpha \right) $ and $N_i^a=N_i^a\left( u^\alpha \right) $ should be
defined by a solution of the Einstein equations (in some models of
Kaluza--Klein gravity \cite{ow} one considers the Einstein--Yang--Mills
fields) for extra dimension gravity.

The metric (\ref{salam}) can be rewritten in a block $(n\times n)\oplus
(m\times m)$ form
\begin{equation}
g_{\alpha \beta }=\left(
\begin{array}{ll}
g_{ij} & 0 \\
0 & h_{ab}
\end{array}
\right)  \label{diagm}
\end{equation}
with respect to some anholonomic frames (N--elongated basis), co--frame
(N--elongated co--basis),
\begin{eqnarray}
\delta _\alpha &=&\frac \delta {\partial u^\alpha }=\left( \delta
_i=\partial _i-N_i^b\partial _b,\delta _a=\partial _a\right) ,  \label{dder}
\\
\delta ^\alpha &=&\delta u^\alpha =\left( \delta ^i=d^i=dx^i,\delta
^a=dy^a+N_i^adx^i\right) ,  \label{ddif}
\end{eqnarray}
which satisfy the anholonomy relations
\[
\delta _\alpha \delta _\beta -\delta _\beta \delta _\alpha =w_{\alpha \beta
}^\gamma \delta _\gamma
\]
with the anholonomy coefficients computed as
\begin{equation}
w_{ij}^k=0,w_{aj}^k=0,w_{ab}^k=0,w_{ab}^c=0,w_{ij}^a=\delta _iN_j^a-\delta
_jN_i^a,w_{ja}^b=-w_{aj}^b=\partial _aN_j^b.  \label{anholonomy}
\end{equation}

In Refs. \cite{sal} the coefficients $N_i^a$ (hereafter, N--coefficients)
were treated as some $U(1),$ $SU(2)$ or $SU(3)$ gauge fields (depending on
the extra dimension $m).$ There are another classes of gravity models which
are constructed on vector (or tangent) bundles generalizing the Finsler
geometry \cite{ma}. In such approaches the set of functions $N_i^a$ were
stated to define a structure of nonlinear connection and the variables $y^a$
were taken to parametrize fibers in some bundles. In the theory of locally
anisotropic (super) strings and supergravity, and gauge generalizations of
the so--called Finsler--Kaluza--Klein gravity the coefficients $N_i^a$ were
suggested to be found from some alternative string models in low energy
limits or from gauge and spinor variants of gravitational field equations
with anholonomic frames and generic local anisotropy \cite{vf}.

The Salam, Strathee and Peracci \cite{sal} idea on a gauge field like status
of the coefficients of off--diagonal metrics in extra dimension gravitity
was developed in a new fashion by applying the method of anholonomic frames
with associated nonlinear connections just on the (pseudo) Riemannian spaces
\cite{v,v2}. The approach allowed to construct new classes of solutions of
Einstein's equations in three (3D), four (4D) and five (5D) dimensions with
generic local anisotropy ({\it e.g.} static black hole and cosmological
solutions with ellipsoidal or toroidal symmetry, various soliton--dilaton 2D
and 3D configurations in 4D gravity, and wormhole and flux tubes with
anisotropic polarizations and/or running on the 5th coordinate constants
with different extensions to backgrounds of rotation ellipsoids, elliptic
cylinders, bipolar and toroidal symmetry and anisotropy).

Recently, it was shown in Refs. \cite{v1} that if we consider off--diagonal
metrics which can be equivalently diagonalized with respect to corresponding
anholonomic frames, the RS theories become substantially locally anisotropic
with variations of constants on extra dimension coordinate or with
anisotropic angular polarizations of effective 4D constants, induced by
higher dimension and/or anholonomic gravitational interactions.

The basic idea on the application of the anholonomic frame method
for constructing exact solutions of the Einstein equations is to
define such N--coefficients when a given type of off--diagonal
metric is diagonalized with respect to some anholonomic frames
(\ref{dder}) and the Einstein equations, re--written in mixed
holonomic and anholonomic variables, trasform into a system of
partial differential equations with separation of variables which
admit exact solutions. This approach differs from the usual
tetradic method where the differential forms and frame bases are
all 'pure' holonomic or 'pure'' anholonomic. In our case the
N--coefficients and associated N--elongated partial derivatives
(\ref{dder}) are chosen as to be some undefined values which at
the final step are fixed as to separate variables and satisfy the
Einstein equations.

The first aim of this paper is to formulate three theorems (and to suggest
the way of their proof) for two off--diagonal metric ansatz which admit
anholonomic transforms resulting in a substantial simplification of the
system of Einstein equations in 5D and 4D gravity. The second aim is to
consider four applications of the anholonomic frame method in order to
construct new classes of exact solutions describing ellipsoidal black holes
with anisotropies and running of constants. We emphasize that is possible to
define classes of warped on the extra dimension coordinate metrics which are
 exact solutions of 5D vacuum gravity. We analyze basic
physical properties of such solutions. We also investigate 5D spacetimes
with anisotropy and cosmological constants.

We use the term 'locally anisotropic' spacetime (or 'anisotropic' spacetime)
for a 5D (4D) pseudo-Riemannian spacetime provided with an anholonomic frame
structure with mixed holonomic and anholonomic variables. The anisotropy of
gravitational interactions is modeled by off--diagonal metrics, or,
equivalently, by theirs diagonalized analogs given with respect to
anholonomic frames.

The paper is organized as follow:\ In Sec. II we formulate three
theorems for two types of off--diagonal metric ansatz, construct
the corresponding exact solutions of 5D vacuum Einstein equations
and illustrate the possibility of extension by introducing matter
fields (the necessary geometric background and some proofs are
presented in the Appendix). We also consider the conditions when
the method generates 4D metrics. In Sec. III we construct two
classes of 5D anisotropic black hole solutions with rotation
ellipsoid horizon and consider  subclasses and reparemetization
of such solutions in order to generate new ones. Sec. IV is
devoted to 4D ellipsoidal black hole solutions. In Sec. V we
extend the method for anisotropic 5D and 4D spacetimes with
cosmological constant, formulate two theorems on basic properties
of the system of field equations and theirs solutions, and give
an example of 5D anisotropic black solution with cosmological
constant. Finally, in Sec. VI, we conclude and discuss the
obtained results.

\section{Off--Diagonal Metrics in Extra Dimension Gravity}

The bulk of solutions of 5D Einstein equations and their
reductions to 4D (like the Schwarzshild solution and brane
generalizations \cite{bh}, metrics with cylindrical and toroidal
symmetry \cite{lemos}, the Friedman--Robertson--Worlker metric
and brane generalizations \cite{wc}) were constructed by using
diagonal metrics and extensions to solutions with rotation, all
given with respect to holonomic coordinate frames of references.
This Section is devoted to a geometrical and nonlinear partial
derivation equations formalism which deals with more general,
generic off--diagonal metrics with respect to coordinate frames,
and anholonomic frames. It summarizes and generalizes various
particular cases and ansatz used for construction of exact
solutions of the Einstein gravitational field equations in 3D, 4D
and 5D gravity \cite{v,v2,v1}.

\subsection{The first ansatz for vacuum Einstein equations}

Let us consider a 5D pseudo--Riemannian spacetime provided with local
coordinates $u^\alpha =(x^i,y^4=v,y^5),$ for $i=1,2,3.$ Our aim is to prove
that a metric ansatz of type (\ref{salam}) can be diagonalized by some
anholonomic transforms with the N--coefficients $N_a^i=N_a^i\left(
x^i,v\right) $ depending on variables $\left( x^i,v\right) $ and to define
the corresponding system of vacuum Einstein equations in the bulk. The exact
solutions of the Einstein equations to be constructed will depend on the
so--called holonomic variables $x^i$ and on one anholonomic (equivalently,
anisotropic) variable $y^4=v.$ In our further considerations every
coordinate from a set $u^\alpha $ can be stated to be time like, 3D space
like or extra dimensional.

For simplicity, the partial derivatives will be denoted like $a^{\times
}=\partial a/\partial x^{1},a^{\bullet }=\partial a/\partial
x^{2},a^{^{\prime }}=\partial a/\partial x^{3},a^{\ast }=\partial a/\partial
v.$

We begin our approach by considering a 5D quadratic line element
\begin{equation}
ds^{2}=g_{\alpha \beta }\left( x^{i},v\right) du^{\alpha }du^{\beta }
\label{metric}
\end{equation}
with the metric coefficients $g_{\alpha \beta }$ parametrized (with respect
to the coordinate frame (\ref{pdif})) by an off--diagonal matrix (ansatz)

{
\begin{equation}
\left[
\begin{array}{ccccc}
g_{1}+w_{1}^{\ 2}h_{4}+n_{1}^{\ 2}h_{5} & w_{1}w_{2}h_{4}+n_{1}n_{2}h_{5} &
w_{1}w_{3}h_{4}+n_{1}n_{3}h_{5} & w_{1}h_{4} & n_{1}h_{5} \\
w_{1}w_{2}h_{4}+n_{1}n_{2}h_{5} & g_{2}+w_{2}^{\ 2}h_{4}+n_{2}^{\ 2}h_{5} &
w_{2}w_{3}h_{4}+n_{2}n_{3}h_{5} & w_{2}h_{4} & n_{2}h_{5} \\
w_{1}w_{3}h_{4}+n_{1}n_{3}h_{5} & w_{2}w_{3}h_{4}+n_{2}n_{3}h_{5} &
g_{3}+w_{3}^{\ 2}h_{4}+n_{3}^{\ 2}h_{5} & w_{3}h_{4} & n_{3}h_{5} \\
w_{1}h_{4} & w_{2}h_{4} & w_{3}h_{4} & h_{4} & 0 \\
n_{1}h_{5} & n_{2}h_{5} & n_{3}h_{5} & 0 & h_{5}
\end{array}
\right] ,  \label{ansatz}
\end{equation}
} where the coefficients are some necessary smoothly class functions of
type:
\begin{eqnarray}
g_{1} &=&\pm 1,g_{2,3}=g_{2,3}(x^{2},x^{3}),h_{4,5}=h_{4,5}(x^{i},v),
\nonumber \\
w_{i} &=&w_{i}(x^{i},v),n_{i}=n_{i}(x^{i},v).  \nonumber
\end{eqnarray}

\begin{lemma}
The quadratic line element (\ref{metric}) with metric coefficients (\ref
{ansatz}) can be diagonalized,
\begin{equation}
\delta
s^{2}=[g_{1}(dx^{1})^{2}+g_{2}(dx^{2})^{2}+g_{3}(dx^{3})^{2}+h_{4}(\delta
v)^{2}+h_{5}(\delta y^{5})^{2}],  \label{dmetric}
\end{equation}
with respect to the anholonomic co--frame $\left( dx^{i},\delta v,\delta
y^{5}\right) ,$ where
\begin{equation}
\delta v=dv+w_{i}dx^{i}\mbox{ and }\delta y^{5}=dy^{5}+n_{i}dx^{i}
\label{ddif1}
\end{equation}
which is dual to the frame $\left( \delta _{i},\partial _{4},\partial
_{5}\right) ,$ where
\begin{equation}
\delta _{i}=\partial _{i}+w_{i}\partial _{4}+n_{i}\partial _{5}.
\label{dder1}
\end{equation}
\end{lemma}

In the Lemma 1 the $N$--coefficients from (\ref{dder}) and (\ref{ddif}) are
parametrized like $N_{i}^{4}=w_{i}$ and $N_{i}^{5}=n_{i}.$

The proof of the Lemma 1 is a trivial computation if we substitute the
values of (\ref{ddif1}) into the quadratic line element (\ref{dmetric}).
Re-writing the metric coefficients with respect to the coodinate basis (\ref
{pdif}) we obtain just the quadratic line element (\ref{metric}) with the
ansatz (\ref{ansatz}).

In the Appendix A we outline the basic formulas from the geometry of
anholonomic frames with mixed holonomic and anholonomic variables and
associated nonlinear connections on (pseudo) Riemannian spaces.

Now we can formulate the

\begin{theorem}
The nontrivial components of the 5D vacuum Einstein equations, $R_{\alpha
}^{\beta }=0,$ (see (\ref{einsteq3}) in the Appendix) for the metric (\ref
{dmetric}) given with respect to anholonomic frames (\ref{ddif1}) and (\ref
{dder1}) are written in a form with separation of variables:
\begin{eqnarray}
R_{2}^{2}=R_{3}^{3}=-\frac{1}{2g_{2}g_{3}}[g_{3}^{\bullet \bullet }-\frac{%
g_{2}^{\bullet }g_{3}^{\bullet }}{2g_{2}}-\frac{(g_{3}^{\bullet })^{2}}{%
2g_{3}}+g_{2}^{^{\prime \prime }}-\frac{g_{2}^{^{\prime }}g_{3}^{^{\prime }}%
}{2g_{3}}-\frac{(g_{2}^{^{\prime }})^{2}}{2g_{2}}] &=&0,  \label{ricci1a} \\
S_{4}^{4}=S_{5}^{5}=-\frac{\beta }{2h_{4}h_{5}} &=&0,  \label{ricci2a} \\
R_{4i}=-w_{i}\frac{\beta }{2h_{5}}-\frac{\alpha _{i}}{2h_{5}} &=&0,
\label{ricci3a} \\
R_{5i}=-\frac{h_{5}}{2h_{4}}\left[ n_{i}^{\ast \ast }+\gamma n_{i}^{\ast }%
\right] &=&0,  \label{ricci4a}
\end{eqnarray}
where
\begin{equation}
\alpha _{i}=\partial _{i}{h_{5}^{\ast }}-h_{5}^{\ast }\partial _{i}\ln \sqrt{%
|h_{4}h_{5}|},\beta =h_{5}^{\ast \ast }-h_{5}^{\ast }[\ln \sqrt{|h_{4}h_{5}|}%
]^{\ast },\gamma =3h_{5}^{\ast }/2h_{5}-h_{4}^{\ast }/h_{4}.  \label{abc}
\end{equation}
\end{theorem}

Here the separation of variables means: 1) we can define a function $%
g_{2}(x^{2},x^{3})$ for a given $g_{3}(x^{2},x^{3}),$ or inversely, to
define a function $g_{2}(x^{2},x^{3})$ for a given $g_{3}(x^{2},x^{3}),$
from equation (\ref{ricci1a}); 2) we can define a function $%
h_{4}(x^{1},x^{2},x^{3},v)$ for a given $h_{5}(x^{1},x^{2},x^{3},v),$ or
inversely, to define a function $h_{5}(x^{1},x^{2},x^{3},v)$ for a given $%
h_{4}(x^{1},x^{2},x^{3},v),$ from equation (\ref{ricci2a}); 3-4) having the
values of $h_{4}$ and $h_{5},$ we can compute the coefficients (\ref{abc})
which allow to solve the algebraic equations (\ref{ricci3a}) and to
integrate two times on $v$ the equations (\ref{ricci4a}) which allow to find
respectively the coefficients $w_{i}(x^{k},v)$ and $n_{i}(x^{k},v).$

The proof of Theorem 1 is a straightforward tensorial and differential
calculus for the components of Ricci tensor (\ref{dricci}) as it is outlined
in the Appendix A. We omit such cumbersome calculations in this paper.

\subsection{The second ansatz for vacuum Einstein equations}

We can consider a generalization of the constructions from the previous
subsection by introducing a conformal factor $\Omega (x^{i},v)$ and
additional deformations of the metric via coefficients $\zeta _{\hat{\imath}%
}(x^{i},v)$ (indices with 'hat' take values like $\hat{{i}}=1,2,3,5).$ The
new metric is written like
\begin{equation}
ds^{2}=\Omega ^{2}(x^{i},v)\hat{{g}}_{\alpha \beta }\left( x^{i},v\right)
du^{\alpha }du^{\beta },  \label{cmetric}
\end{equation}
were the coefficients $\hat{{g}}_{\alpha \beta }$ are parametrized by the
ansatz {\scriptsize
\begin{equation}
\left[
\begin{array}{ccccc}
g_{1}+(w_{1}^{\ 2}+\zeta _{1}^{\ 2})h_{4}+n_{1}^{\ 2}h_{5} &
(w_{1}w_{2}+\zeta _{1}\zeta _{2})h_{4}+n_{1}n_{2}h_{5} & (w_{1}w_{3}+\zeta
_{1}\zeta _{3})h_{4}+n_{1}n_{3}h_{5} & (w_{1}+\zeta _{1})h_{4} & n_{1}h_{5}
\\
(w_{1}w_{2}+\zeta _{1}\zeta _{2})h_{4}+n_{1}n_{2}h_{5} & g_{2}+(w_{2}^{\
2}+\zeta _{2}^{\ 2})h_{4}+n_{2}^{\ 2}h_{5} & (w_{2}w_{3}++\zeta _{2}\zeta
_{3})h_{4}+n_{2}n_{3}h_{5} & (w_{2}+\zeta _{2})h_{4} & n_{2}h_{5} \\
(w_{1}w_{3}+\zeta _{1}\zeta _{3})h_{4}+n_{1}n_{3}h_{5} & (w_{2}w_{3}++\zeta
_{2}\zeta _{3})h_{4}+n_{2}n_{3}h_{5} & g_{3}+(w_{3}^{\ 2}+\zeta _{3}^{\
2})h_{4}+n_{3}^{\ 2}h_{5} & (w_{3}+\zeta _{3})h_{4} & n_{3}h_{5} \\
(w_{1}+\zeta _{1})h_{4} & (w_{2}+\zeta _{2})h_{4} & (w_{3}+\zeta _{3})h_{4}
& h_{4} & 0 \\
n_{1}h_{5} & n_{2}h_{5} & n_{3}h_{5} & 0 & h_{5}+\zeta _{5}h_{4}
\end{array}
\right] .  \label{ansatzc}
\end{equation}
}

Such 5D pseudo--Riemannian metrics are considered to have second order
anisotropy \cite{vf,ma}. For trivial values $\Omega =1$ and $\zeta _{\hat{%
\imath}}=0,$ the squared line interval (\ref{cmetric}) transforms into (\ref
{metric}).

\begin{lemma}
The quadratic line element (\ref{cmetric}) with metric coefficients (\ref
{ansatzc}) can be diagonalized,
\begin{equation}
\delta s^{2}=\Omega
^{2}(x^{i},v)[g_{1}(dx^{1})^{2}+g_{2}(dx^{2})^{2}+g_{3}(dx^{3})^{2}+h_{4}(%
\hat{{\delta }}v)^{2}+h_{5}(\delta y^{5})^{2}],  \label{cdmetric}
\end{equation}
with respect to the anholonomic co--frame $\left( dx^{i},\hat{{\delta }}%
v,\delta y^{5}\right) ,$ where
\begin{equation}
\delta v=dv+(w_{i}+\zeta _{i})dx^{i}+\zeta _{5}\delta y^{5}\mbox{ and }%
\delta y^{5}=dy^{5}+n_{i}dx^{i}  \label{ddif2}
\end{equation}
which is dual to the frame $\left( \hat{{\delta }}_{i},\partial _{4},\hat{{%
\partial }}_{5}\right) ,$ where
\begin{equation}
\hat{{\delta }}_{i}=\partial _{i}-(w_{i}+\zeta _{i})\partial
_{4}+n_{i}\partial _{5},\hat{{\partial }}_{5}=\partial _{5}-\zeta
_{5}\partial _{4}.  \label{dder2}
\end{equation}
\end{lemma}

In the Lemma 2 the $N$--coefficients from (\ref{pder}) and (\ref{dder}) are
parametrized in the first order anisotropy (with three anholonomic, $x^{i},$
and two anholonomic, $y^{4}$ and $y^{5},$ coordinates) like $N_{i}^{4}=w_{i}$
and $N_{i}^{5}=n_{i}$ and in the second order anisotropy (on the second
'shell', \ with four anholonomic, $(x^{i},y^{5}),$ and one anholonomic,$%
y^{4},$ coordinates) with $N_{\hat{{i}}}^{5}=\zeta _{\hat{{i}}},$ in this
work we state, for symplicity, $\zeta _{\hat{{i}}}=0.$

The Theorem 1 can be extended as to include the generalization to the second
ansatz:

\begin{theorem}
The nontrivial components of the 5D vacuum Einstein equations, $R_{\alpha
}^{\beta }=0,$ (see (\ref{einsteq3}) in the Appendix) for the metric (\ref
{cdmetric}) given with respect to anholonomic frames (\ref{ddif2}) and (\ref
{dder2}) are written in \ the same form as in the system (\ref{ricci1a})--(%
\ref{ricci4a}) with the additional conditions that
\begin{equation}
\hat{{\delta }}_{i}h_{4}=0\mbox{\ and\  }\hat{{\delta }}_{i}\Omega =0
\label{conf1}
\end{equation}
and the values $\zeta _{\hat{{i}}}=\left( \zeta _{{i}},\zeta _{{5}}=0\right)
$ are found as to be a unique solution of (\ref{conf1}); for instance, if
\begin{equation}
\Omega ^{q_{1}/q_{2}}=h_{4}~(q_{1}\mbox{ and }q_{2}\mbox{ are integers}),
\label{confq}
\end{equation}
$\zeta _{{i}}$ satisfy the equations \
\begin{equation}
\partial _{i}\Omega -(w_{i}+\zeta _{{i}})\Omega ^{\ast }=0.  \label{confeq}
\end{equation}
\end{theorem}

The proof of Theorem 2 consists from a straightforward calculation of the
components of the Ricci tensor (\ref{dricci}) as it is outlined in the
Appendix. The simplest way is to use the calculus for Theorem 1 and then to
compute deformations of the canonical d--connection (\ref{dcon}). \ Such
deformations induce corresponding deformations of the Ricci tensor (\ref
{dricci}). \ The condition that we have the same values of the Ricci tensor
for the (\ref{ansatz}) and (\ref{ansatzc}) results in equations (\ref{conf1}%
) and (\ref{confeq}) which are compatible, for instance, if $\Omega
^{q_{1}/q_{2}}=h_{4}.$\ There are also another possibilities to satisfy the
condition (\ref{conf1}), for instance, if $\Omega =\Omega _{1}$ $\Omega
_{2}, $ we can consider that $h_{4}=\Omega _{1}^{q_{1}/q_{2}}$ $\Omega
_{2}^{q_{3}/q_{4}}$ $\ $for some integers $q_{1},q_{2},q_{3}$ and $q_{4}.$

\subsection{General solutions}

The surprising result is that we are able to construct exact solutions of
the 5D vacuum Einstein equations for both types of the ansatz (\ref{ansatz})
and (\ref{ansatzc}):

\begin{theorem}
The system of second order nonlinear partial differential equations (\ref
{ricci1a})--(\ref{ricci4a}) and (\ref{confeq}) can be solved in general form
if there are given some values of functions $g_{2}(x^{2},x^{3})$ (or $%
g_{3}(x^{2},x^{3})),h_{4}\left( x^{i},v\right) $ (or $h_{5}\left(
x^{i},v\right) )$ and $\Omega \left( x^{i},v\right) :$

\begin{itemize}
\item  The general solution of equation (\ref{ricci1a}) can be written in
the form
\begin{equation}
\varpi =g_{[0]}\exp [a_{2}\widetilde{x}^{2}\left( x^{2},x^{3}\right) +a_{3}%
\widetilde{x}^{3}\left( x^{2},x^{3}\right) ],  \label{solricci1a}
\end{equation}
were $g_{[0]},a_{2}$ and $a_{3}$ are some constants and the functions $%
\widetilde{x}^{2,3}\left( x^{2},x^{3}\right) $ define coordinate transforms $%
x^{2,3}\rightarrow \widetilde{x}^{2,3}$ for which the 2D line element
becomes conformally flat, i. e.
\begin{equation}
g_{2}(x^{2},x^{3})(dx^{2})^{2}+g_{3}(x^{2},x^{3})(dx^{3})^{2}\rightarrow
\varpi \left[ (d\widetilde{x}^{2})^{2}+\epsilon (d\widetilde{x}^{3})^{2}%
\right] .  \label{con10}
\end{equation}

\item  The equation (\ref{ricci2a}) relates two functions $h_{4}\left(
x^{i},v\right) $ and $h_{5}\left( x^{i},v\right) $. There are two
possibilities:

a) to compute
\begin{eqnarray}
\sqrt{|h_{5}|} &=&h_{5[1]}\left( x^{i}\right) +h_{5[2]}\left( x^{i}\right)
\int \sqrt{|h_{4}\left( x^{i},v\right) |}dv,~h_{4}^{\ast }\left(
x^{i},v\right) \neq 0;  \nonumber \\
&=&h_{5[1]}\left( x^{i}\right) +h_{5[2]}\left( x^{i}\right) v,h_{4}^{\ast
}\left( x^{i},v\right) =0,  \label{p2}
\end{eqnarray}
for some functions $h_{5[1,2]}\left( x^{i}\right) $ stated by boundary
conditions;

b) or, inversely, to compute $h_{4}$ for a given $h_{5}\left( x^{i},v\right)
,h_{5}^{\ast }\neq 0,$%
\begin{equation}
\sqrt{|h_{4}|}=h_{[0]}\left( x^{i}\right) (\sqrt{|h_{5}\left( x^{i},v\right)
|})^{\ast },  \label{p1}
\end{equation}
with $h_{[0]}\left( x^{i}\right) $ given by boundary conditions.

\item  The exact solutions of (\ref{ricci3a}) for $\beta \neq 0$ is
\begin{equation}
w_{k}=\partial _{k}\ln [\sqrt{|h_{4}h_{5}|}/|h_{5}^{\ast }|]/\partial
_{v}\ln [\sqrt{|h_{4}h_{5}|}/|h_{5}^{\ast }|],  \label{w}
\end{equation}
with $\partial _{v}=\partial /\partial v$ and $h_{5}^{\ast }\neq 0.$ If $%
h_{5}^{\ast }=0,$ or even $h_{5}^{\ast }\neq 0$ but $\beta =0,$ the
coefficients $w_{k}$ could be arbitrary functions on $\left( x^{i},v\right)
. $ \ For vacuum Einstein equations this is a degenerated case which imposes
the the compatibility conditions $\beta =\alpha _{i}=0,$ which are
satisfied, for instance, if the $h_{4}$ and $h_{5}$ are related as in the
formula (\ref{p1}) but with $h_{[0]}\left( x^{i}\right) =const.$

\item  The exact solution of (\ref{ricci4a}) is
\begin{eqnarray}
n_{k} &=&n_{k[1]}\left( x^{i}\right) +n_{k[2]}\left( x^{i}\right) \int
[h_{4}/(\sqrt{|h_{5}|})^{3}]dv,~h_{5}^{\ast }\neq 0;  \nonumber \\
&=&n_{k[1]}\left( x^{i}\right) +n_{k[2]}\left( x^{i}\right) \int
h_{4}dv,\qquad ~h_{5}^{\ast }=0;  \label{n} \\
&=&n_{k[1]}\left( x^{i}\right) +n_{k[2]}\left( x^{i}\right) \int [1/(\sqrt{%
|h_{5}|})^{3}]dv,~h_{4}^{\ast }=0,  \nonumber
\end{eqnarray}
for some functions $n_{k[1,2]}\left( x^{i}\right) $ stated by boundary
conditions.

\item  The exact solution of (\ref{confeq}) is given by some arbitrary
functions $\zeta _{i}=\zeta _{i}\left( x^{i},v\right) $ if \ both $\partial
_{i}\Omega =0$ and $\Omega ^{\ast }=0,$ we chose $\zeta _{i}=0$ for $\Omega
=const,$ and
\begin{eqnarray}
\zeta _{i} &=&-w_{i}+(\Omega ^{\ast })^{-1}\partial _{i}\Omega ,\quad \Omega
^{\ast }\neq 0,  \label{confsol} \\
&=&(\Omega ^{\ast })^{-1}\partial _{i}\Omega ,\quad \Omega ^{\ast }\neq 0,%
\mbox{ for vacuum solutions}.  \nonumber
\end{eqnarray}
\end{itemize}
\end{theorem}

\bigskip We note that a transform (\ref{con10}) is always possible for 2D
metrics and the explicit form of solutions depends on chosen system of 2D
coordinates and on the signature $\epsilon =\pm 1.$ In the simplest case the
equation (\ref{ricci1a}) is solved by arbitrary two functions $g_{2}(x^{3})$
and $g_{3}(x^{2}).$ The equation (\ref{ricci2a}) is satisfied by arbitrary
pairs of coefficients $h_{4}\left( x^{i},v\right) $ and $h_{5[0]}\left(
x^{i}\right) .$

The proof of Theorem 3 is given in the Appendix B.

\subsection{Consequences of Theorems 1--3}

We consider three important consequences of the Lemmas and Theorems
formulated in this Section:

\begin{corollary}
The non--trivial diagonal components of the Einstein tensor, $G_{\beta
}^{\alpha }=R_{\beta }^{\alpha }-\frac{1}{2}R\quad \delta _{\beta }^{\alpha
},$ for the metric (\ref{dmetric}), given with respect to anholonomic
N--bases, are
\begin{equation}
G_{1}^{1}=-\left( R_{2}^{2}+S_{4}^{4}\right)
,G_{2}^{2}=G_{3}^{3}=-S_{4}^{4},G_{4}^{4}=G_{5}^{5}=-R_{2}^{2}.
\label{einstdiag}
\end{equation}
So, the dynamics of the system is defined by two values $R_{2}^{2}$ and $%
S_{4}^{4}.$ The rest of non--diagonal components of the Ricci (Einstein
tensor) are compensated by fixing corresponding values of N--coefficients.
\end{corollary}

The formulas (\ref{einstdiag}) are obtained following the relations for the
Ricci tensor (\ref{ricci1a})--(\ref{ricci4a}).

\begin{corollary}
We can extend the system of 5D vacuum Einstein equations (\ref{ricci1a})--(%
\ref{ricci4a}) by introducing matter fields for which the energy--momentum
tensor $\Upsilon _{\alpha \beta }$ given with respect to anholonomic frames
satisfy the conditions
\begin{equation}
\Upsilon _{1}^{1}=\Upsilon _{2}^{2}+\Upsilon _{4}^{4},\Upsilon
_{2}^{2}=\Upsilon _{3}^{3},\Upsilon _{4}^{4}=\Upsilon _{5}^{5}.
\label{emcond}
\end{equation}
\end{corollary}

We note that, in general, the tensor $\Upsilon _{\alpha \beta }$ for the
non--vacuum Einstein equations,
\[
R_{\alpha \beta }-\frac{1}{2}g_{\alpha \beta }R=\kappa \Upsilon _{\alpha
\beta },
\]
is not symmetric because with respect to anholonomic frames there are
imposed constraints which makes non symmetric the Ricci and Einstein tensors
(the symmetry conditions hold only with respect to holonomic, coordinate
frames; for details see the Appendix and the formulas (\ref{einsteq2})).

For simplicity, in our further investigations we shall consider only
diagonal matter sources, given with respect to anholonomic frames,
satisfying the conditions
\begin{equation}
\kappa \Upsilon _{2}^{2}=\kappa \Upsilon _{3}^{3}=\Upsilon _{2},\kappa
\Upsilon _{4}^{4}=\kappa \Upsilon _{5}^{5}=\Upsilon _{4},\mbox{ and }%
\Upsilon _{1}=\Upsilon _{2}+\Upsilon _{4},  \label{diagemt}
\end{equation}
where $\kappa $ is the gravitational coupling constant. In this case the
equations (\ref{ricci1a}) and (\ref{ricci2a}) are respectively generalized
to
\begin{equation}
R_{2}^{2}=R_{3}^{3}=-\frac{1}{2g_{2}g_{3}}[g_{3}^{\bullet \bullet }-\frac{%
g_{2}^{\bullet }g_{3}^{\bullet }}{2g_{2}}-\frac{(g_{3}^{\bullet })^{2}}{%
2g_{3}}+g_{2}^{^{\prime \prime }}-\frac{g_{2}^{^{\prime }}g_{3}^{^{\prime }}%
}{2g_{3}}-\frac{(g_{2}^{^{\prime }})^{2}}{2g_{2}}]=-\Upsilon _{4}
\label{ricci1b}
\end{equation}
and
\begin{equation}
S_{4}^{4}=S_{5}^{5}=-\frac{\beta }{2h_{4}h_{5}}=-\Upsilon _{2}.
\label{ricci2b}
\end{equation}

\begin{corollary}
The class of metrics (\ref{cmetric}) satisfying vacuum Einstein equations (%
\ref{ricci1a})--(\ref{ricci4a}) and (\ref{confeq}) contains as particular
cases some solutions when the Schwarzschild potential $\Phi =-M/(M_{{\rm p}%
}^{2}r)$, where $M_{{\rm p}}$ is the effective Planck mass on the brane, is
modified to
\[
\Phi =-{\frac{M\sigma _{m}}{M_{{\rm p}}^{2}r}}+{\frac{Q\sigma _{q}}{2r^{2}}}%
\,,
\]
where the `tidal charge' parameter $Q$ may be positive or negative.
\end{corollary}

As proofs of this corollary we can consider the Refs \cite{v1} where the
possibility to modify anisotropically the Newton law via effective
anisotropic masses $M\sigma _{m},$ or by anisotropic effective 4D Plank
constants, renormalized like $\sigma _{m}/M_{{\rm p}}^{2},$ and with
''effective'' electric charge, $Q\sigma _{q}$ was recently emphasized (see
also the end of Section III in this paper). For diagonal metrics, in the
locally isotropic limit, we put the effective polarizations $\sigma
_{m}=\sigma _{q}=1.$

\subsection{Reduction from 5D to 4D gravity}

The above presented results are for generic off--diagonal metrics of
gravitational fields, anholonomic transforms and nonlinear field equations.
Reductions to a lower dimensional theory are not trivial in such cases. We
give a detailed analysis of this procedure.

The symplest way to construct a $5D\rightarrow 4D$ reduction for the ansatz (%
\ref{ansatz}) and (\ref{ansatzc}) is to eliminate from formulas the variable
$x^{1}$ and to consider a 4D space (parametrized by local coordinates $%
\left( x^{2},x^{3},v,y^{5}\right) )$ being trivially embedded into 5D space
(parametrized by local coordinates $\left( x^{1},x^{2},x^{3},v,y^{5}\right) $
with $g_{11}=\pm 1,g_{1\underline{\alpha }}=0,\underline{\alpha }=2,3,4,5)$
with further possible conformal and anholonomic transforms depending only on
variables $\left( x^{2},x^{3},v\right) .$ We admit that the 4D metric $g_{%
\underline{\alpha }\underline{\beta }}$ could be of arbitrary signature. In
order to emphasize that some coordinates are stated just for a such 4D space
we underline the Greek indices, $\underline{\alpha },\underline{\beta },...$
\ and the Latin indices from the meadle of alphabet, $\underline{i},%
\underline{j},...=2,3,$ where $u^{\underline{\alpha }}=\left( x^{\underline{i%
}},y^{a}\right) =\left( x^{2},x^{3},y^{4},y^{5}\right) .$

In result, the analogs of Lemmas 1and 2, Theorems 1-3 and Corollaries 1-3
can be reformulated for 4D gravity with mixed holonomic--anholonomic
variables. We outline here the most important properties of a such reduction.

\begin{itemize}
\item  The line element (\ref{metric}) with ansatz (\ref{ansatz}) and the
line element (\ref{metric}) with (\ref{ansatzc}) are respectively
transformed on 4D space to the values:

The first type 4D quadratic line element is taken
\begin{equation}
ds^{2}=g_{\underline{\alpha }\underline{\beta }}\left( x^{\underline{i}%
},v\right) du^{\underline{\alpha }}du^{\underline{\beta }}  \label{metric4}
\end{equation}
with the metric coefficients $g_{\alpha \beta }$ parametrized (with respect
to the coordinate frame (\ref{pdif}) in 4D) by an off--diagonal matrix
(ansatz)

{
\begin{equation}
\left[
\begin{array}{cccc}
g_{2}+w_{2}^{\ 2}h_{4}+n_{2}^{\ 2}h_{5} & w_{2}w_{3}h_{4}+n_{2}n_{3}h_{5} &
w_{2}h_{4} & n_{2}h_{5} \\
w_{2}w_{3}h_{4}+n_{2}n_{3}h_{5} & g_{3}+w_{3}^{\ 2}h_{4}+n_{3}^{\ 2}h_{5} &
w_{3}h_{4} & n_{3}h_{5} \\
w_{2}h_{4} & w_{3}h_{4} & h_{4} & 0 \\
n_{2}h_{5} & n_{3}h_{5} & 0 & h_{5}
\end{array}
\right] ,  \label{ansatz4}
\end{equation}
} where the coefficients are some necessary smoothly class functions of
type:
\begin{eqnarray}
g_{2,3} &=&g_{2,3}(x^{2},x^{3}),h_{4,5}=h_{4,5}(x^{\underline{k}},v),
\nonumber \\
w_{\underline{i}} &=&w_{\underline{i}}(x^{\underline{k}},v),n_{\underline{i}%
}=n_{\underline{i}}(x^{\underline{k}},v);~\underline{i},\underline{k}=2,3.
\nonumber
\end{eqnarray}

The anholonomically and conformally transformed 4D line element is
\begin{equation}
ds^{2}=\Omega ^{2}(x^{\underline{i}},v)\hat{{g}}_{\underline{\alpha }%
\underline{\beta }}\left( x^{\underline{i}},v\right) du^{\underline{\alpha }%
}du^{\underline{\beta }},  \label{cmetric4}
\end{equation}
were the coefficients $\hat{{g}}_{\underline{\alpha }\underline{\beta }}$
are parametrized by the ansatz {\scriptsize
\begin{equation}
\left[
\begin{array}{cccc}
g_{2}+(w_{2}^{\ 2}+\zeta _{2}^{\ 2})h_{4}+n_{2}^{\ 2}h_{5} &
(w_{2}w_{3}++\zeta _{2}\zeta _{3})h_{4}+n_{2}n_{3}h_{5} & (w_{2}+\zeta
_{2})h_{4} & n_{2}h_{5} \\
(w_{2}w_{3}++\zeta _{2}\zeta _{3})h_{4}+n_{2}n_{3}h_{5} & g_{3}+(w_{3}^{\
2}+\zeta _{3}^{\ 2})h_{4}+n_{3}^{\ 2}h_{5} & (w_{3}+\zeta _{3})h_{4} &
n_{3}h_{5} \\
(w_{2}+\zeta _{2})h_{4} & (w_{3}+\zeta _{3})h_{4} & h_{4} & 0 \\
n_{2}h_{5} & n_{3}h_{5} & 0 & h_{5}+\zeta _{5}h_{4}
\end{array}
\right] .  \label{ansatzc4}
\end{equation}
}where $\zeta _{\underline{i}}=\zeta _{\underline{i}}\left( x^{\underline{k}%
},v\right) $ and we shall restrict our considerations for $\zeta _{5}=0.$

\item  In the 4D analog of Lemma 1 we have
\begin{equation}
\delta s^{2}=[g_{2}(dx^{2})^{2}+g_{3}(dx^{3})^{2}+h_{4}(\delta
v)^{2}+h_{5}(\delta y^{5})^{2}],  \label{dmetric4}
\end{equation}
with respect to the anholonomic co--frame $\left( dx^{\underline{i}},\delta
v,\delta y^{5}\right) ,$ where
\begin{equation}
\delta v=dv+w_{\underline{i}}dx^{\underline{i}}\mbox{ and }\delta
y^{5}=dy^{5}+n_{\underline{i}}dx^{\underline{i}}  \label{ddif4}
\end{equation}
which is dual to the frame $\left( \delta _{\underline{i}},\partial
_{4},\partial _{5}\right) ,$ where
\begin{equation}
\delta _{\underline{i}}=\partial _{\underline{i}}+w_{\underline{i}}\partial
_{4}+n_{\underline{i}}\partial _{5}.  \label{dder4}
\end{equation}

\item  In the conditions of the 4D variant of Theorem 1 we have the same
equations (\ref{ricci1a})--(\ref{ricci4a}) were we must put $%
h_{4}=h_{4}\left( x^{\underline{k}},v\right) $ and $h_{5}=h_{5}\left( x^{%
\underline{k}},v\right) .$ As a consequence we have that $\alpha _{i}\left(
x^{k},v\right) \rightarrow \alpha _{\underline{i}}\left( x^{\underline{k}%
},v\right) ,\beta =\beta \left( x^{\underline{k}},v\right) $ and $\gamma
=\gamma \left( x^{\underline{k}},v\right) $ which result that $w_{\underline{%
i}}=w_{\underline{i}}\left( x^{\underline{k}},v\right) $ and $n_{\underline{i%
}}=n_{\underline{i}}\left( x^{\underline{k}},v\right) .$

\item  The respective formulas from Lemma 2, for $\zeta _{5}=0,$ transform
into
\begin{equation}
\delta s^{2}=\Omega ^{2}(x^{\underline{i}%
},v)[g_{2}(dx^{2})^{2}+g_{3}(dx^{3})^{2}+h_{4}(\hat{{\delta }}%
v)^{2}+h_{5}(\delta y^{5})^{2}],  \label{cdmetric4}
\end{equation}
with respect to the anholonomic co--frame $\left( dx^{\underline{i}},\hat{{%
\delta }}v,\delta y^{5}\right) ,$ where
\begin{equation}
\delta v=dv+(w_{\underline{i}}+\zeta _{\underline{i}})dx^{\underline{i}}%
\mbox{ and }\delta y^{5}=dy^{5}+n_{\underline{i}}dx^{\underline{i}}
\label{ddif24}
\end{equation}
which is dual to the frame $\left( \hat{{\delta }}_{\underline{i}},\partial
_{4},\hat{{\partial }}_{5}\right) ,$ where
\begin{equation}
\hat{{\delta }}_{\underline{i}}=\partial _{\underline{i}}-(w_{\underline{i}%
}+\zeta _{\underline{i}})\partial _{4}+n_{\underline{i}}\partial _{5},\hat{{%
\partial }}_{5}=\partial _{5}.  \label{dder24}
\end{equation}

\item  The formulas (\ref{conf1}) and (\ref{confeq}) from Theorem 2 must be
modified into a 4D form
\begin{equation}
\hat{{\delta }}_{\underline{i}}h_{4}=0\mbox{\ and\  }\hat{{\delta }}_{%
\underline{i}}\Omega =0  \label{conf14}
\end{equation}
and the values $\zeta _{\hat{{i}}}=\left( \zeta \underline{_{{i}}},\zeta _{{5%
}}=0\right) $ are found as to be a unique solution of (\ref{conf1}); for
instance, if
\[
\Omega ^{q_{1}/q_{2}}=h_{4}~(q_{1}\mbox{ and }q_{2}\mbox{ are integers}),
\]
$\zeta _{\underline{{i}}}$ satisfy the equations \
\begin{equation}
\partial _{\underline{i}}\Omega -(w_{\underline{i}}+\zeta \underline{_{{i}}}%
)\Omega ^{\ast }=0.  \label{confeq4}
\end{equation}

\item  One holds the same formulas (\ref{p2})-(\ref{n}) from the Theorem 3
on the general form of exact solutions with that difference that their 4D
analogs are to be obtained by reductions of holonomic indices, $\underline{i}%
\rightarrow i,$ and holonomic coordinates, $x^{i}\rightarrow x^{\underline{i}%
},$ i. e. in the 4D solutions there is not contained the variable $x^{1}.$

\item  The formulae (\ref{einstdiag}) for the nontrivial coefficients of the
Einstein tensor in 4D stated by the Corollary 1 are \ written
\begin{equation}
G_{2}^{2}=G_{3}^{3}=-S_{4}^{4},G_{4}^{4}=G_{5}^{5}=-R_{2}^{2}.
\label{einstdiag4}
\end{equation}

\item  For symmetries of the Einstein tensor (\ref{einstdiag4}) \ we can
introduce a matter field source with a diagonal energy momentum tensor,
like\ it is stated in the Corollary 2 by the conditions (\ref{emcond}),
which in 4D are transformed into
\begin{equation}
\Upsilon _{2}^{2}=\Upsilon _{3}^{3},\Upsilon _{4}^{4}=\Upsilon _{5}^{5}.
\label{emcond4}
\end{equation}

\item  In 4D Einstein gravity we are not having violations of the Newton law
as it was state in Corollary 3 for 5D. Nevertheless, off--diagonal and
anholonomic frames can induce an anholonomic \ particle and field dynamics,
for instance, with deformations of horizons of black holes, which can be
modeled by an effective anisotropic renormalization of constants if some
conditions are satisfied \cite{v,v2}.
\end{itemize}

There were constructed and analyzed various classes of exact
solutions of the Einstein equations (both in the vacuum, reducing
to the system (\ref {ricci1a}), (\ref{ricci2a}), (\ref{ricci3a})
and (\ref{ricci4a}) and non--vacuum, reducing to (\ref{ricci1b}),
(\ref{ricci2b}), (\ref{ricci3a}) and (\ref{ricci4a}), cases) in
3D, 4D and 5D gravity \cite{v,v1}. The aim of the next Sections
III -- V is to prove that such solutions contain warped factors
which in the vacuum case are induced by a second order anisotropy.
We shall analyze some classes of such exact solutions with
running constants and/or their anisotropic polarizations induced
from extra dimension gravitational interactions.

\section{5D Ellipsoidal Black Holes}

Our goal is to apply the anholonomic frame method as to construct such exact
solutions of vacuum 5D Einstein equations as they will be static ones but,
for instance, with ellipsoidal horizon for a diagonal metric given with
respect to some well defined anholonomic frames. If such metrics are
redefined with respect to usual coordinate frames, they are described by
some particular cases of off--diagonal ansatz of type (\ref{ansatz}), or (%
\ref{ansatzc}) which results in a very sophysticate form of the Einstein
equations. That why it was not possible to construct such solutions in the
past, before elaboration of the anholonomic frame method with associated
nonlinear connection structure which allows to find exact solutions of the
Einstein equations for very general off--diagonal metric ansatz.

By using anholonomic transforms the Schwarzschild and Reissner-N\"{o}rdstrom
solutions were generalized in anisotropic forms with deformed horizons,
anisoropic polarizations and running constants both in the Einstein and
extra dimension gravity (see Refs. \cite{v,v1}). It was shown that there are
possible anisotropic solutions which preserve the local Lorentz symmetry.
and that at large radial distances from the horizon the anisotropic
configurations transform into the usual one with spherical symmetry. So, the
solutions with anisotropic rotation ellipsoidal horizons do not contradict
the well known Israel and Carter theorems \cite{israel} which were proved in
the assumption of spherical symmetry at asymptotics. The vacuum metrics
presented here differ from anisotropic black hole solutions investigated in
Refs. \cite{v,v1}.

\subsection{The Schwarzschild solution in ellipsoidal coordinates}

Let us consider the system of {\it \ isotropic spherical coordinates} $(\rho
,\theta ,\varphi ),$ \thinspace where the isotropic radial coordinate $\rho $
is related with the usual radial coordinate $r$ via the relation $r=\rho
\left( 1+r_{g}/4\rho \right) ^{2}$ for $r_{g}=2G_{[4]}m_{0}/c^{2}$ being the
4D gravitational radius of a point particle of mass $m_{0},$ $%
G_{[4]}=1/M_{P[4]}^{2}$ is the 4D Newton constant expressed via Plank mass $%
M_{P[4]}$ (following modern string/brane theories, $M_{P[4]}$ can be
considered as a value induced from extra dimensions). We put the light speed
constant $c=1.$ This system of coordinates is considered for the so--called
isotropic representation of the Schwarzschild solution \cite{ll}
\begin{equation}
dS^{2}=\left( \frac{\widehat{\rho }-1}{\widehat{\rho }+1}\right)
^{2}dt^{2}-\rho _{g}^{2}\left( \frac{\widehat{\rho }+1}{\widehat{\rho }}%
\right) ^{4}\left( d\widehat{\rho }^{2}+\widehat{\rho }^{2}d\theta ^{2}+%
\widehat{\rho }^{2}\sin ^{2}\theta d\varphi ^{2}\right) ,  \label{schw}
\end{equation}
where, for our further considerations, we re--scaled the isotropic radial
coordinate as $\widehat{\rho }=\rho /\rho _{g},$ with $\rho _{g}=r_{g}/4.$
The metric (\ref{schw}) is a vacuum static solution of 4D Einstein equations
with spherical symmetry describing the gravitational field of a point
particle of mass $m_{0}.$ It has a singularity for $r=0$ and a spherical
horizon for $r=r_{g},$ or, in re--scaled isotropic coordinates, for $%
\widehat{\rho }=1.$ We emphasize that this solution is parametrized by a
diagonal metric given with respect to holonomic coordinate frames.

We also introduce the {\it \ rotation ellipsoid coordinates} (in our case
considered as alternatives to the isotropic radial coordinates) \cite{korn} $%
(u,\lambda ,\varphi )$ with $0\leq u<\infty ,0\leq \lambda \leq \pi ,0\leq
\varphi \leq 2\pi ,$ where $\sigma =\cosh u\geq 1$ are related with the
isotropic 3D Cartezian coordinates
\begin{equation}
(\tilde{x}=\widetilde{\rho }\sinh u\sin \lambda \cos \varphi ,\tilde{y}=%
\widetilde{\rho }\sinh u\sin \lambda \sin \varphi ,\tilde{z}=\widetilde{\rho
}\cosh u\cos \lambda )  \label{rec}
\end{equation}
and define an elongated rotation ellipsoid hypersurface
\begin{equation}
\left( \tilde{x}^{2}+\tilde{y}^{2}\right) /(\sigma ^{2}-1)+\tilde{z}%
^{2}/\sigma ^{2}=\widetilde{\rho }^{2}. \label{reh}
\end{equation}
with $\sigma =\cosh u.$ The 3D metric on a such hypersurface is
\[
dS_{(3D)}^{2}=g_{u u}du^{2}+g_{\lambda \lambda }d\lambda
^{2}+g_{\varphi \varphi }d\varphi ^{2},
\]
where
\[
g_{u u}=g_{\lambda \lambda }=\widetilde{\rho }^{2}\left( \sinh
^{2}u+\sin ^{2}\lambda \right) ,g_{\varphi \varphi
}=\widetilde{\rho }^{2}\sinh ^{2}u\sin ^{2}\lambda .
\]

We can relate the rotation ellipsoid coordinates $\left( u,\lambda ,\varphi
\right) $ from (\ref{rec}) with the isotropic radial coordinates $\left(
\widehat{\rho },\theta ,\varphi \right) $, scaled by the constant $\rho
_{g}, $ from (\ref{schw}) as
\[
\widetilde{\rho }=1,\sigma =\cosh u=\widehat{\rho }
\]
and deform the Schwarzschild metric by introducing ellipsoidal
coordinates and a new  horizon defined by the condition that
vanishing of the metric coefficient before $dt^2$ describe an
elongated rotation ellipsoid hypersurface (\ref{reh}),
\begin{eqnarray}
dS_{(S)}^{2} &=&\left( \frac{\cosh u-1}{\cosh u+1}\right) ^{2}dt^{2}-\rho
_{g}^{2}\left( \frac{\cosh u+1}{\cosh u}\right) ^{4}(\sinh ^{2}u+\sin
^{2}\lambda )  \label{schel} \\
&&\times \lbrack du^{2}+d\lambda ^{2}+\frac{\sinh ^{2}u~\sin ^{2}\lambda }{%
\sinh ^{2}u+\sin ^{2}\lambda }d\varphi ^{2}].  \nonumber
\end{eqnarray}
The ellipsoidally deformed metric (\ref{schel}) does not satisfy
the vacuum Einstein equations, but at long distances from the
horizon it transforms into the usual Schwarzchild solution
(\ref{schw}).

For our further considerations we introduce two Classes (A and B)
of 4D auxiliary pseudo--Riemannian metrics, also given in
ellipsoid coordinates, being some conformal transforms of
(\ref{schel}), like
\[
dS_{(S)}^{2}=\Omega _{A,B}\left( u,\lambda \right) dS_{(A,B)}^{2}
\]
but which are not supposed to be solutions of the Einstein equations:

\begin{itemize}
\item  Metric of Class A:
\begin{equation}
dS_{(A)}^{2}=-du^{2}-d\lambda ^{2}+a(u,\lambda )d\varphi ^{2}+b(u,\lambda
)dt^{2}],  \label{auxm1}
\end{equation}
where
\[
a(u,\lambda )=-\frac{\sinh ^{2}u~\sin ^{2}\lambda }{\sinh ^{2}u+\sin
^{2}\lambda }\mbox{ and }b(u,\lambda )=-\frac{(\cosh u-1)^{2}\cosh ^{4}u}{%
\rho _{g}^{2}(\cosh u+1)^{6}(\sinh ^{2}u+\sin ^{2}\lambda )},
\]
which results in the metric (\ref{schel}) by multiplication on the conformal
factor
\begin{equation}
\Omega _{A}\left( u,\lambda \right) =\rho _{g}^{2}\frac{(\cosh u+1)^{4}}{%
\cosh ^{4}u}(\sinh ^{2}u+\sin ^{2}\lambda ).  \label{auxm1c}
\end{equation}

\item  Metric of Class B:
\begin{equation}
dS^{2}=g(u,\lambda )\left( du^{2}+d\lambda ^{2}\right) -d\varphi
^{2}+f(u,\lambda )dt^{2},  \label{auxm2}
\end{equation}
where
\[
g(u,\lambda )=-\frac{\sinh ^{2}u+\sin ^{2}\lambda }{\sinh ^{2}u~\sin
^{2}\lambda }\mbox{ and }f(u,\lambda )=\frac{(\cosh u-1)^{2}\cosh ^{4}u}{%
\rho _{g}^{2}(\cosh u+1)^{6}\sinh ^{2}u\sin ^{2}\lambda },
\]
which results in the metric (\ref{schel}) by multiplication on the conformal
factor
\[
\Omega _{B}\left( u,\lambda \right) =\rho _{g}^{2}\frac{(\cosh u+1)^{4}}{%
\cosh ^{4}u}\sinh ^{2}u\sin ^{2}\lambda .
\]
\end{itemize}

Now it is possible to generate exact solutions of the Einstein equations
with rotation ellipsoid horizons and anisotropic polarizations and running
of constants by performing corresponding anholonomic transforms as the
solutions will have an horizon parametrized by a hypersurface like rotation
ellipsoid and gravitational (extra dimensional or nonlinear 4D)
renormalization of the constant $\rho _{g}$ of the Schwarzschild solution, $%
\rho _{g}\rightarrow \overline{\rho }_{g}=\omega \rho _{g},$ where the
dependence of the function $\omega $ on some holonomic or anholonomic
coordinates depend on the type of anisotropy. For some solutions we can
treat $\omega $ as a factor modeling running of the gravitational constant,
induced, induced from extra dimension, in another cases we may consider $%
\omega $ as a nonlinear gravitational polarization which model some
anisotropic distributions of masses and matter fields and/or anholonomic
vacuum gravitational interactions.

\subsection{Ellipsoidal 5D metrics of Class A}

In this subsection we consider four classes of 5D vacuum solutions which are
related to the metric of Class A (\ref{auxm1}) and to the Schwarzshild
metric in ellipsoidal coordinates (\ref{schel}).

Let us parametrize the 5D coordinates as $\left( x^{1}=\chi
,x^{2}=u,x^{3}=\lambda ,y^{4}=v,y^{5}=p\right) ,$ where the solutions with
the so--called $\varphi $--anisotropy will be constructed for $\left(
v=\varphi ,p=t\right) $ and the solutions with $t$--anisotropy will be
stated for $\left( v=t,p=\varphi \right) $ (in brief, we shall write $\ $\
respective $\varphi $--solutions and $t$--solutions).

\subsubsection{Class A solutions with ansatz (\ref{ansatz}):}

We take an off--diagonal metric ansatz of type (\ref{ansatz}) (equivalently,
(\ref{metric})) by reprezenting
\[
g_{1}=\pm 1,g_{2}=-1,g_{3}=-1,h_{4}=\eta _{4}(x^{i},v)h_{4(0)}(x^{i})%
\mbox{
and }h_{5}=\eta _{5}(x^{i},v)h_{5(0)}(x^{i}),
\]
where $\eta _{4,5}(x^{i},v)$ are corresponding ''gravitational
renormalizations'' of the metric coefficients $h_{4,5(0)}(x^{i}).$ For $%
\varphi $--solutions we state $h_{4(0)}=a(u,\lambda )$ and $%
h_{5(0)}=b(u,\lambda )$ (inversely, for \ $t$--solutions, $%
h_{4(0)}=b(u,\lambda )$ and $h_{5(0)}=a(u,\lambda )).$ \

Next we consider a renormalized gravitational 'constant' $\overline{\rho }%
_{g}=\omega \rho _{g},$ were for $\varphi $--solutions\ the receptivity $%
\omega =\omega \left( x^{i},v\right) $ is included in the graviational
polarization $\eta _{5}$ as $\eta _{5}=\left[ \omega \left( x^{i},\varphi
\right) \right] ^{-2},$ or for $t$--solutions is included in $\eta _{4},$
when $\eta _{4}=\left[ \omega \left( x^{i},t\right) \right] ^{-2}.$ We can
construct an exact solution of the 5D vacuum Einstein equations if, for
explicit dependencies on anisotropic coordinate, the metric coefficients $%
h_{4}$ and $h_{5}$ are related by formula (\ref{p1}) with $h_{[0]}\left(
x^{i}\right) =h_{(0)}=const$ (see the Theorem 3, with statements on formulas
(\ref{p1}) and (\ref{w})), which in its turn imposes a corresponding
relation between $\eta _{4}$ and $\eta _{5},$%
\[
\eta _{4}h_{4(0)}(x^{i})=h_{(0)}^{2}h_{5(0)}(x^{i})\left[ \left( \sqrt{|\eta
_{5}|}\right) ^{\ast }\right] ^{2}.
\]
In result, we express the polarizations $\eta _{4}$ and $\eta _{5}$ via the
value of receptivity $\omega ,$
\begin{equation}
\eta _{4}\left( \chi ,u,\lambda ,\varphi \right) =h_{(0)}^{2}\frac{%
b(u,\lambda )}{a(u,\lambda )}\left\{ \left[ \omega ^{-1}\left( \chi
,u,\lambda ,\varphi \right) \right] ^{\ast }\right\} ^{2},\eta _{5}\left(
\chi ,u,\lambda ,\varphi \right) =\omega ^{-2}\left( \chi ,u,\lambda
,\varphi \right) ,  \label{etap}
\end{equation}
for $\varphi $--solutions , and
\begin{equation}
\eta _{4}\left( \chi ,u,\lambda ,t\right) =\omega ^{-2}\left( \chi
,u,\lambda ,t\right) ,\eta _{5}\left( \chi ,u,\lambda ,t\right) =h_{(0)}^{-2}%
\frac{b(u,\lambda )}{a(u,\lambda )}\left[ \int dt\omega ^{-1}\left( \chi
,u,\lambda ,t\right) \right] ^{2},  \label{etat}
\end{equation}
for $t$--solutions, where $a(u,\lambda )$ and $b(u,\lambda )$ are those from
(\ref{auxm1}).

For vacuum configurations, following the discussions of formula (\ref{w}) in
Theorem 3, we put $w_{i}=0.$ The next step is to find the values of $n_{i}$
by introducing $h_{4}=\eta _{4}h_{4(0)}$ and $h_{5}=\eta _{5}h_{5(0)}$ into
the formula \ (\ref{n}), which, for convenience, is expressed via general
coefficients $\eta _{4}$ and $\eta _{5},$ with the functions $n_{k[2]}\left(
x^{i}\right) $ redefined as to contain the values $h_{(0)}^{2},$ $%
a(u,\lambda )$ and $b(u,\lambda )$
\begin{eqnarray}
n_{k} &=&n_{k[1]}\left( x^{i}\right) +n_{k[2]}\left( x^{i}\right) \int [\eta
_{4}/(\sqrt{|\eta _{5}|})^{3}]dv,~\eta _{5}^{\ast }\neq 0;  \label{nel} \\
&=&n_{k[1]}\left( x^{i}\right) +n_{k[2]}\left( x^{i}\right) \int \eta
_{4}dv,\qquad ~\eta _{5}^{\ast }=0;  \nonumber \\
&=&n_{k[1]}\left( x^{i}\right) +n_{k[2]}\left( x^{i}\right) \int [1/(\sqrt{%
|\eta _{5}|})^{3}]dv,~\eta _{4}^{\ast }=0.  \nonumber
\end{eqnarray}

\bigskip By introducing the formulas (\ref{etap}) for $\varphi $--solutions
(or (\ref{etat}) for $t$--solutions) and fixing some boundary condition, in
order to state the values of coefficients $n_{k[1,2]}\left( x^{i}\right) $
we can express the ansatz components $n_{k}\left( x^{i},\varphi \right) $ as
integrals of some functions of $\omega \left( x^{i},\varphi \right) $ and $%
\partial _{\varphi }\omega \left( x^{i},\varphi \right) $ (or, we can
express the ansatz components $n_{k}\left( x^{i},t\right) $ as integrals of
some functions of $\omega \left( x^{i},t\right) $ and $\partial _{t}\omega
\left( x^{i},t\right) ).$ We do not present an explicit form of such
formulas because they depend on the type of receptivity $\omega =\omega
\left( x^{i},v\right) ,$ which must be defined experimentally, or from some
quantum models of gravity in the quasi classical limit. We preserved a
general dependence on coordinates $x^{i}$ which reflect the fact that there
is a freedom in fixing holonomic coordinates (for instance, on ellipsoidal
hypersurface and its extensions to 4D and 5D spacetimes). \ For simplicity,
we write that $n_{i}$ are some functionals of $\{x^{i},\omega \left(
x^{i},v\right) ,\omega ^{\ast }\left( x^{i},v\right) \}$
\[
n_{i}\{x,\omega ,\omega ^{\ast }\}=n_{i}\{x^{i},\omega \left( x^{i},v\right)
,\omega ^{\ast }\left( x^{i},v\right) \}.
\]

In conclusion, we constructed two exact solutions of the 5D vacuum Einstein
equations, defined by the ansatz (\ref{ansatz}) with coordinates and
coefficients stated by the data:
\begin{eqnarray}
\mbox{$\varphi$--solutions} &:&(x^{1}=\chi ,x^{2}=u,x^{3}=\lambda
,y^{4}=v=\varphi ,y^{5}=p=t),g_{1}=\pm 1,  \nonumber \\
g_{2} &=&-1,g_{3}=-1,h_{4(0)}=a(u,\lambda ),h_{5(0)}=b(u,\lambda ),%
\mbox{see
(\ref{auxm1})};  \nonumber \\
h_{4} &=&\eta _{4}(x^{i},\varphi )h_{4(0)}(x^{i}),h_{5}=\eta
_{5}(x^{i},\varphi )h_{5(0)}(x^{i}),  \nonumber \\
\eta _{4} &=&h_{(0)}^{2}\frac{b(u,\lambda )}{a(u,\lambda )}\left\{ \left[
\omega ^{-1}\left( \chi ,u,\lambda ,\varphi \right) \right] ^{\ast }\right\}
^{2},\eta _{5}=\omega ^{-2}\left( \chi ,u,\lambda ,\varphi \right) ,
\nonumber \\
w_{i} &=&0,n_{i}\{x,\omega ,\omega ^{\ast }\}=n_{i}\{x^{i},\omega \left(
x^{i},\varphi \right) ,\omega ^{\ast }\left( x^{i},\varphi \right) \}.
\label{sol5p1}
\end{eqnarray}
and
\begin{eqnarray}
\mbox{$t$--solutions} &:&(x^{1}=\chi ,x^{2}=u,x^{3}=\lambda
,y^{4}=v=t,y^{5}=p=\varphi ),g_{1}=\pm 1,  \nonumber \\
g_{2} &=&-1,g_{3}=-1,h_{4(0)}=b(u,\lambda ),h_{5(0)}=a(u,\lambda ),%
\mbox{see
(\ref{auxm1})};  \nonumber \\
h_{4} &=&\eta _{4}(x^{i},t)h_{4(0)}(x^{i}),h_{5}=\eta
_{5}(x^{i},t)h_{5(0)}(x^{i}),  \nonumber \\
\eta _{4} &=&\omega ^{-2}\left( \chi ,u,\lambda ,t\right) ,\eta
_{5}=h_{(0)}^{-2}\frac{b(u,\lambda )}{a(u,\lambda )}\left[ \int dt~\omega
^{-1}\left( \chi ,u,\lambda ,t\right) \right] ^{2},  \nonumber \\
w_{i} &=&0,n_{i}\{x,\omega ,\omega ^{\ast }\}=n_{i}\{x^{i},\omega \left(
x^{i},t\right) ,\omega ^{\ast }\left( x^{i},t\right) \}.  \label{sol5t1}
\end{eqnarray}

Both types of solutions have a horizon parametrized by a rotation ellipsoid
hypersurface (as the condition of vanishing of \ the ''time'' metric
coefficient states, i. e. when the function $b(u,\lambda )=0)$. $\ $These
solutions are generically anholonomic (anisotropic) because in the locally
isotropic limit, when $\eta _{4},\eta _{5},$ $\omega \rightarrow 1$ and $%
n_{i}\rightarrow 0,$ they reduce to the coefficients of the
metric (\ref {auxm1}). The last one is not an exact solution of
4D vacuum Einstein equations, but it is a conformal transform of
the 4D Schwarzschild solution with a further trivial extension to
5D. With respect to the anholonomic frames adapted to the
coefficients $n_{i}$ (see (\ref{ddif1})), the obtained solutions
have diagonal metric coefficients being very similar to the
Schwarzschild metric (\ref{schel}) written in ellipsoidal
coordinates. We can treat such solutions as black hole ones with
a point particle mass put in one of the focuses of rotation
ellipsoid hypersurface (for flattened ellipsoids the mass should
be placed on the circle described by ellipse's focuses under
rotation; we omit such details in this work which were presented
for 4D gravity in Ref. \cite{v}).

 The initial data for anholonomic frames and the chosen
configuration of gravitational interactions in the bulk lead to
deformed ''ellipsoidal'' horizons even for static configurations.
The solutions admit anisotropic polarizations on ellipsoidal and
angular coordinates $\left( u,\lambda \right) $ and running of
constants on time $t$ and/or on extra dimension coordinate $\chi
$. Such renormalizations of constants are defined by the nonlinear
configuration of the 5D vacuum gravitational field and depend on
introduced receptivity function $\omega \left( x^{i},v\right) $
which is to be considered an intrinsic characteristics of the 5D
vacuum gravitational 'ether', emphasizing the possibility \ of
nonlinear self--polarization of gravitational fields.

Finally, we note that the data (\ref{sol5p1}) and (\ref{sol5t1}) parametrize
two very different classes of solutions. The first one is for static 5D
vacuum black hole configurations with explicit dependence on anholonomic
coordinate $\varphi $ and possible renormalizations on the rest of 3D space
coordinates $u$ and $\lambda $ and on the 5th coordinate $\chi .$ The second
class of solutions are similar to the static solutions but with an
emphasized anholonomic time running of constants and with possible
anisotropic dependencies on coordinates $(u,\lambda ,\chi ).$

\subsubsection{Class A solutions with ansatz (\ref{ansatzc}):}

We construct here 5D vacuum $\varphi $-- and $t$--solutions parametrized by
an ansatz with conformal factor $\Omega (x^{i},v)$ (see (\ref{ansatzc}) and (%
\ref{cdmetric})). Let us consider conformal factors parametrized as $\Omega
=\Omega _{\lbrack 0]}(x^{i})\Omega _{\lbrack 1]}(x^{i},v).$ We can generate
from the data (\ref{sol5p1}) (or (\ref{sol5t1})) an exact solution of vacuum
Einstein equations if there are satisfied the conditions (\ref{confq}) and (%
\ref{confsol}), i. e.
\[
\Omega _{\lbrack 0]}^{q_{1}/q_{2}}\Omega _{\lbrack 1]}^{q_{1}/q_{2}}=\eta
_{4}h_{4(0)},
\]
for some integers $q_{1}$ and $q_{2},$ and there are defined the second
anisotropy coefficients
\[
\zeta _{i}=\left( \partial _{i}\ln |\Omega _{\lbrack 0]}\right) |)~\left(
\ln |\Omega _{\lbrack 1]}|\right) ^{\ast }+\left( \Omega _{\lbrack 1]}^{\ast
}\right) ^{-1}\partial _{i}\Omega _{\lbrack 1]}.
\]
So, taking a $\varphi $-- or $t$--solution with corresponding values of $%
h_{4}=\eta _{4}h_{4(0)},$\ for some $q_{1}$ and $q_{2},$ we obtain new exact
solutions, called in brief, $\varphi _{c}$-- or $t_{c}$--solutions (with the
index ''c'' pointing to an ansatz with conformal factor), of the vacuum 5D
Einstein equations given in explicit form by the data:
\begin{eqnarray}
\mbox{$\varphi_c$--solutions} &:&(x^{1}=\chi ,x^{2}=u,x^{3}=\lambda
,y^{4}=v=\varphi ,y^{5}=p=t),g_{1}=\pm 1,  \nonumber \\
g_{2} &=&-1,g_{3}=-1,h_{4(0)}=a(u,\lambda ),h_{5(0)}=b(u,\lambda ),%
\mbox{see
(\ref{auxm1})};  \nonumber \\
h_{4} &=&\eta _{4}(x^{i},\varphi )h_{4(0)}(x^{i}),h_{5}=\eta
_{5}(x^{i},\varphi )h_{5(0)}(x^{i}),  \nonumber \\
\eta _{4} &=&h_{(0)}^{2}\frac{b(u,\lambda )}{a(u,\lambda )}\left\{ \left[
\omega ^{-1}\left( \chi ,u,\lambda ,\varphi \right) \right] ^{\ast }\right\}
^{2},\eta _{5}=\omega ^{-2}\left( \chi ,u,\lambda ,\varphi \right) ,
\nonumber \\
w_{i} &=&0,n_{i}\{x,\omega ,\omega ^{\ast }\}=n_{i}\{x^{i},\omega \left(
x^{i},\varphi \right) ,\omega ^{\ast }\left( x^{i},\varphi \right) \},\Omega
=\Omega _{\lbrack 0]}(x^{i})\Omega _{\lbrack 1]}(x^{i},\varphi )
\label{sol5pc} \\
\zeta _{i} &=&\left( \partial _{i}\ln |\Omega _{\lbrack 0]}\right) |)~\left(
\ln |\Omega _{\lbrack 1]}|\right) ^{\ast }+\left( \Omega _{\lbrack 1]}^{\ast
}\right) ^{-1}\partial _{i}\Omega _{\lbrack 1]},\eta _{4}a=\Omega _{\lbrack
0]}^{q_{1}/q_{2}}(x^{i})\Omega _{\lbrack 1]}^{q_{1}/q_{2}}(x^{i},\varphi ).
\nonumber
\end{eqnarray}
and
\begin{eqnarray}
\mbox{$t_c$--solutions} &:&(x^{1}=\chi ,x^{2}=u,x^{3}=\lambda
,y^{4}=v=t,y^{5}=p=\varphi ),g_{1}=\pm 1,  \nonumber \\
g_{2} &=&-1,g_{3}=-1,h_{4(0)}=b(u,\lambda ),h_{5(0)}=a(u,\lambda ),%
\mbox{see
(\ref{auxm1})};  \nonumber \\
h_{4} &=&\eta _{4}(x^{i},t)h_{4(0)}(x^{i}),h_{5}=\eta
_{5}(x^{i},t)h_{5(0)}(x^{i}),  \nonumber \\
\eta _{4} &=&\omega ^{-2}\left( \chi ,u,\lambda ,t\right) ,\eta
_{5}=h_{(0)}^{-2}\frac{b(u,\lambda )}{a(u,\lambda )}\left[ \int dt~\omega
^{-1}\left( \chi ,u,\lambda ,t\right) \right] ^{2},  \nonumber \\
w_{i} &=&0,n_{i}\{x,\omega ,\omega ^{\ast }\}=n_{i}\{x^{i},\omega \left(
x^{i},t\right) ,\omega ^{\ast }\left( x^{i},t\right) \},\Omega =\Omega
_{\lbrack 0]}(x^{i})\Omega _{\lbrack 1]}(x^{i},t)  \label{sol5tc} \\
\zeta _{i} &=&\left( \partial _{i}\ln |\Omega _{\lbrack 0]}\right) |)~\left(
\ln |\Omega _{\lbrack 1]}|\right) ^{\ast }+\left( \Omega _{\lbrack 1]}^{\ast
}\right) ^{-1}\partial _{i}\Omega _{\lbrack 1]},\eta _{4}a=\Omega _{\lbrack
0]}^{q_{1}/q_{2}}(x^{i})\Omega _{\lbrack 1]}^{q_{1}/q_{2}}(x^{i},t).
\nonumber
\end{eqnarray}

These solutions have two very interesting properties: 1) they admit a warped
factor on the 5th coordinate, like $\Omega _{\lbrack 1]}^{q_{1}/q_{2}}\sim
\exp [-k|\chi |],$ which in our case is constructed for an anisotropic 5D
vacuum gravitational configuration and not following a brane configuration
like in Refs. \cite{rs}; 2) we can impose such conditions on the receptivity
$\omega \left( x^{i},v\right) $ as to obtain in the locally isotropic limit
just the Schwarzschild metric (\ref{schel}) trivially embedded into the 5D
spacetime.

Let us analyze the second property in details. We have to chose the
conformal factor as to be satisfied three conditions:
\begin{equation}
\Omega _{\lbrack 0]}^{q_{1}/q_{2}}=\Omega _{A},\Omega _{\lbrack
1]}^{q_{1}/q_{2}}\eta _{4}=1,\Omega _{\lbrack 1]}^{q_{1}/q_{2}}\eta _{5}=1,
\label{cond1a}
\end{equation}
were $\Omega _{A}$ is that from (\ref{confq}). The last two conditions are
possible if
\begin{equation}
\eta _{4}^{-q_{1}/q_{2}}\eta _{5}=1,  \label{cond2}
\end{equation}
which selects a specific form of receptivity $\omega \left( x^{i},v\right) .$
\ Putting into (\ref{cond2}) the values $\eta _{4}$ and $\eta _{5}$
respectively from (\ref{sol5pc}), or (\ref{sol5tc}), we obtain some
differential, or integral, relations of the unknown $\omega \left(
x^{i},v\right) ,$ which results that
\begin{eqnarray}
\omega \left( x^{i},\varphi \right) &=&\left( 1-q_{1}/q_{2}\right)
^{-1-q_{1}/q_{2}}\left[ h_{(0)}^{-1}\sqrt{|a/b|}\varphi +\omega _{\lbrack
0]}\left( x^{i}\right) \right] ,\mbox{ for }\varphi _{c}\mbox{--solutions};
\label{cond1} \\
\omega \left( x^{i},t\right) &=&\left[ \left( q_{1}/q_{2}-1\right) h_{(0)}%
\sqrt{|a/b|}t+\omega _{\lbrack 1]}\left( x^{i}\right) \right]
^{1-q_{1}/q_{2}},\mbox{ for }t_{c}\mbox{--solutions},  \nonumber
\end{eqnarray}
for some arbitrary functions $\omega _{\lbrack 0]}\left( x^{i}\right) $ and $%
\omega _{\lbrack 1]}\left( x^{i}\right) .$ \ So, receptivities of particular
form like (\ref{cond1}) allow us to obtain in the locally isotropic limit
just the Schwarzschild metric.

We conclude this subsection by \ the remark: the \ vacuum 5D metrics solving
the Einstein equations describe a nonlinear gravitational dynamics which
under some particular boundary conditions and parametrizations of metric's
coefficients can model anisotropic solutions transforming, in a
corresponding locally isotropic limit, in some well known exact solutions
like Schwarzschild, Reissner-N\"{o}rdstrom, Taub NUT, various type of
wormhole, solitonic and disk solutions (see details in Refs. \cite{v,v2,v1}%
). Here we emphasize that, in general, an anisotropic solution (parametrized
by an off--diagonal ansatz) could not have a locally isotropic limit to a
diagonal metric with respect to some holonomic coordinate frames. By some
boundary conditions and suggested type of horizons, singularities,
symmetries and topological configuration such solutions model new classes of
black hole/tori, wormholes and another type of solutions which defines a
generic anholonomic gravitational field dynamics and has not locally
isotropic limits.

\subsection{Ellipsiodal 5D metrics of Class B}

In this subsection we construct and analyze another two classes of 5D vacuum
solutions which are related to the metric of Class B (\ref{auxm2}) and which
can be reduced to the Schwarzshild metric in ellipsoidal coordinates (\ref
{schel}) by corresponding parametrizations of receptivity $\omega \left(
x^{i},v\right) $. We emphasize that because the function $g(u,\lambda )$
from (\ref{auxm2}) is not a solution of equation (\ref{ricci1a}) we
introduce an auxiliary factor $\varpi $ $(u,\lambda )$ for which $\varpi g$
becames a such solution, then we consider conformal factors parametrized as $%
\Omega =\varpi ^{-1}$ $\Omega _{\lbrack 2]}\left( x^{i},v\right) $ and find
solutions parametrized by the ansatz (\ref{ansatzc}) and anholonomic metric
interval (\ref{cdmetric}).

Besause the method of definition of such solutions is similar to that from
previous subsection, in our further considerations we shall omit intermediar
computations and present directly the data which select the respective
configurations for $\varphi _{c}$--solutions and $t_{c}$--solutions.

The Class B of 5D solutions with conformal factor are parametrized by the
data:

\begin{eqnarray}
\mbox{$\varphi_c$--solutions} &:&(x^{1}=\chi ,x^{2}=u,x^{3}=\lambda
,y^{4}=v=\varphi ,y^{5}=p=t),g_{1}=\pm 1,  \nonumber \\
g_{2} &=&g_{3}=\varpi (u,\lambda )g(u,\lambda ),h_{4(0)}=-\varpi (u,\lambda
),h_{5(0)}=\varpi (u,\lambda )f(u,\lambda ),\mbox{see
(\ref{auxm2})};  \nonumber \\
\varpi &=&g^{-1}\varpi _{0}\exp [a_{2}u+a_{3}\lambda ],~\varpi
_{0},a_{2},a_{3}=const;~\mbox{see
(\ref{solricci1a})}  \nonumber \\
h_{4} &=&\eta _{4}(x^{i},\varphi )h_{4(0)}(x^{i}),h_{5}=\eta
_{5}(x^{i},\varphi )h_{5(0)}(x^{i}),  \nonumber \\
\eta _{4} &=&-h_{(0)}^{2}f(u,\lambda )\left\{ \left[ \omega ^{-1}\left( \chi
,u,\lambda ,\varphi \right) \right] ^{\ast }\right\} ^{2},\eta _{5}=\omega
^{-2}\left( \chi ,u,\lambda ,\varphi \right) ,  \label{sol5p} \\
w_{i} &=&0,n_{i}\{x,\omega ,\omega ^{\ast }\}=n_{i}\{x^{i},\omega \left(
x^{i},\varphi \right) ,\omega ^{\ast }\left( x^{i},\varphi \right) \},\Omega
=\varpi ^{-1}(u,\lambda )\Omega _{\lbrack 2]}(x^{i},\varphi )  \nonumber \\
\zeta _{i} &=&\partial _{i}\ln |\varpi |)~\left( \ln |\Omega _{\lbrack
2]}|\right) ^{\ast }+\left( \Omega _{\lbrack 2]}^{\ast }\right)
^{-1}\partial _{i}\Omega _{\lbrack 2]},\eta _{4}=-\varpi
^{-q_{1}/q_{2}}(x^{i})\Omega _{\lbrack 2]}^{q_{1}/q_{2}}(x^{i},\varphi ).
\nonumber
\end{eqnarray}
and
\begin{eqnarray}
\mbox{$t_c$--solutions} &:&(x^{1}=\chi ,x^{2}=u,x^{3}=\lambda
,y^{4}=v=t,y^{5}=p=\varphi ),g_{1}=\pm 1,  \nonumber \\
g_{2} &=&g_{3}=\varpi (u,\lambda )g(u,\lambda ),h_{4(0)}=\varpi (u,\lambda
)f(u,\lambda ),h_{5(0)}=-\varpi (u,\lambda ),\mbox{see
(\ref{auxm2})};  \nonumber \\
\varpi &=&g^{-1}\varpi _{0}\exp [a_{2}u+a_{3}\lambda ],~\varpi
_{0},a_{2},a_{3}=const,~\mbox{see
(\ref{solricci1a})}  \nonumber \\
h_{4} &=&\eta _{4}(x^{i},t)h_{4(0)}(x^{i}),h_{5}=\eta
_{5}(x^{i},t)h_{5(0)}(x^{i}),  \nonumber \\
\eta _{4} &=&\omega ^{-2}\left( \chi ,u,\lambda ,t\right) ,\eta
_{5}=-h_{(0)}^{-2}f(u,\lambda )\left[ \int dt~\omega ^{-1}\left( \chi
,u,\lambda ,t\right) \right] ^{2},  \label{sol5t} \\
w_{i} &=&0,n_{i}\{x,\omega ,\omega ^{\ast }\}=n_{i}\{x^{i},\omega \left(
x^{i},t\right) ,\omega ^{\ast }\left( x^{i},t\right) \},\Omega =\varpi
^{-1}(u,\lambda )\Omega _{\lbrack 2]}(x^{i},t)  \nonumber \\
\zeta _{i} &=&\partial _{i}(\ln |\varpi |)~\left( \ln |\Omega _{\lbrack
2]}|\right) ^{\ast }+\left( \Omega _{\lbrack 2]}^{\ast }\right)
^{-1}\partial _{i}\Omega _{\lbrack 2]},\eta _{4}=-\varpi
^{-q_{1}/q_{2}}(x^{i})\Omega _{\lbrack 2]}^{q_{1}/q_{2}}(x^{i},t).  \nonumber
\end{eqnarray}
where the coeffiecients $n_{i}$ can be found explicitly by introducing the
corresponding values $\eta _{4}$ and $\eta _{5}$ in formula (\ref{nel}).

By a procedure similar to the solutions of Class A (see previous subsection)
we can find the conditions when the solutions (\ref{sol5p}) and (\ref{sol5t}%
) will have in the locally anisotropic limit the Schwarzshild solutions,
which impose corresponding parametrizations and dependencies on $\Omega
_{\lbrack 2]}(x^{i},v)$ and $\omega \left( x^{i},v\right) $ like (\ref
{cond1a}) and (\ref{cond1}). We omit these formulas because, in general, for
aholonomic configurations and nonlinear solutions there are not hard
arguments to prefer any holonomic limits of such off--diagonal metrics.

Finally, in this Section, we remark that for the considered
classes of ellipsoidal black hole solutions the so--called
$tt$--components of metric contain modifications of the
Schwarzschild potential
\[
\Phi =-\frac{M}{M_{P[4]}^{2}r}\mbox{ into }\Phi =-\frac{M\omega \left(
x^{i},v\right) }{M_{P[4]}^{2}r},
\]
where $M_{P[4]}$ is the usual 4D Plank constant, and this is given with
respect to the corresponding aholonomic frame of reference. The receptivity $%
\omega \left( x^{i},v\right) $ could model corrections warped on extra
dimension coordinate, $\chi ,$ which for our solutions are induced by
anholonomic vacuum gravitational interactions in the bulk and not from a
brane configuration in $AdS_{5}$ spacetime. In the vacuum case $k$ is a
constant which chareacterizes the receptivity for bulk vacuum gravitational
polarizations.

\section{4D Ellipsoidal Black Holes}

For the ansatz (\ref{ansatz4}), without conformal factor, some
classes of ellipsoidal solutions of 4D Einstein equations were
constructed in Ref. \cite {v} with further generalizations and
applications to brane physics \cite{v1} . The goal of this
Section is to consider some alternative variants, both with and
without conformal factors and for different coordinate
parametrizations and types of anisotropies. The bulk of 5D
solutions from the previous Section are reduced into
corresponding 4D ones if one eliminates the 5th coordinate $\chi $
from\ the formulas and  the off--diagonal ansatz (\ref{ansatz4})
and (\ref {ansatzc4}) are considered.

\subsection{Ellipsiodal 5D metrics of Class A}

Let us parametrize the 4D coordinates as $(x^{\underline{i}},y^{a})=\left(
x^{2}=u,x^{3}=\lambda ,y^{4}=v,y^{5}=p\right) ;$ for the $\varphi $%
--solutions we shall take $\left( v=\varphi ,p=t\right) $ and for
the solutions $t$--solutions we shall consider $\left(
v=t,p=\varphi \right) $. Following the prescription from
subsection IIE we can write down the data for solutions without
proofs and computations.

\subsubsection{Class A solutions with ansat (\ref{ansatz4}):}

The off--diagonal metric ansatz of type (\ref{ansatz4}) (equivalently, (\ref
{metric})) \ with the data
\begin{eqnarray}
\mbox{$\varphi$--solutions} &:&(x^{2}=u,x^{3}=\lambda ,y^{4}=v=\varphi
,y^{5}=p=t)  \nonumber \\
g_{2} &=&-1,g_{3}=-1,h_{4(0)}=a(u,\lambda ),h_{5(0)}=b(u,\lambda ),%
\mbox{see
(\ref{auxm1})};  \nonumber \\
h_{4} &=&\eta _{4}(u,\lambda ,\varphi )h_{4(0)}(u,\lambda ),h_{5}=\eta
_{5}(u,\lambda ,\varphi )h_{5(0)}(u,\lambda ),  \nonumber \\
\eta _{4} &=&h_{(0)}^{2}\frac{b(u,\lambda )}{a(u,\lambda )}\left\{ \left[
\omega ^{-1}\left( u,\lambda ,\varphi \right) \right] ^{\ast }\right\}
^{2},\eta _{5}=\omega ^{-2}\left( u,\lambda ,\varphi \right) ,  \nonumber \\
w_{\underline{i}} &=&0,n_{\underline{}i}\{x,\omega ,\omega ^{\ast
}\}=n_{i}\{u,\lambda ,\omega \left( u,\lambda ,\varphi \right) ,\omega
^{\ast }\left( u,\lambda ,\varphi \right) \}.  \label{sol4p1}
\end{eqnarray}
and
\begin{eqnarray}
\mbox{$t$--solutions} &:&(x^{2}=u,x^{3}=\lambda ,y^{4}=v=t,y^{5}=p=\varphi )
\nonumber \\
g_{2} &=&-1,g_{3}=-1,h_{4(0)}=b(u,\lambda ),h_{5(0)}=a(u,\lambda ),%
\mbox{see
(\ref{auxm1})};  \nonumber \\
h_{4} &=&\eta _{4}(u,\lambda ,t)h_{4(0)}(u,\lambda ),h_{5}=\eta
_{5}(u,\lambda ,t)h_{5(0)}(u,\lambda ),  \nonumber \\
\eta _{4} &=&\omega ^{-2}\left( u,\lambda ,t\right) ,\eta _{5}=h_{(0)}^{-2}%
\frac{b(u,\lambda )}{a(u,\lambda )}\left[ \int dt~\omega ^{-1}\left(
u,\lambda ,t\right) \right] ^{2},  \nonumber \\
w_{\underline{i}} &=&0,n_{\underline{i}}\{x,\omega ,\omega ^{\ast }\}=n_{%
\underline{i}}\{u,\lambda ,\omega \left( u,\lambda ,t\right) ,\omega ^{\ast
}\left( u,\lambda ,t\right) \}.  \label{sol4t1}
\end{eqnarray}
where the $n_{\underline{i}}$ are computed
\begin{eqnarray}
n_{\underline{k}} &=&n_{\underline{k}[1]}\left( u,\lambda \right)
+n_{\underline{k}[2]}\left( u,\lambda \right) \int [\eta
_{4}/(\sqrt{|\eta _{5}|})^{3}]dv,~\eta _{5}^{\ast }\neq 0;
\label{nem4} \\
&=&n_{\underline{k}[1]}\left( u,\lambda \right)
+n_{\underline{k}[2]}\left( u,\lambda \right) \int
\eta _{4}dv,\qquad ~\eta _{5}^{\ast }=0;  \nonumber \\
&=&n_{\underline{k}[1]}\left( u,\lambda \right)
+n_{\underline{k}[2]}\left( u,\lambda \right) \int
[1/(\sqrt{|\eta _{5}|})^{3}]dv,~\eta _{4}^{\ast }=0.  \nonumber
\end{eqnarray}
 These solutions have the same ellipsoidal
symmetries and properties stated for their 5D analogs
(\ref{sol5p1}) and for (\ref{sol5t1}) with that difference that
there are not any warped factors and extra dimension
dependencies. We emphasize that the solutions defined by the
formulas (\ref{sol4p1}) and (\ref{sol4t1}) do not result in a
locally isotropic limit into an exact solution having diagonal
coefficients with respect to some holonomic coordinate frames.
The data introduced in this subsection are for generic 4D vacuum
solutions of the Einstein equations parametrized by off--diagonal
metrics. The renormalization of constants and metric coefficients
have a 4D nonlinear vacuum gravitational origin and reflects a
corresponding anholonomic dynamics.

\subsubsection{Class A solutions with ansatz (\ref{ansatzc4}):}

The 4D vacuum $\varphi $-- and $t$--solutions parametrized by an ansatz with
conformal factor $\Omega (u,\lambda ,v)$ (see (\ref{ansatzc4}) and (\ref
{cdmetric4})). Let us consider conformal factors parametrized as $\Omega
=\Omega _{\lbrack 0]}(u,\lambda )\Omega _{\lbrack 1]}(u,\lambda ,v).$ The
data are
\begin{eqnarray}
\mbox{$\varphi_c$--solutions} &:&(x^{2}=u,x^{3}=\lambda ,y^{4}=v=\varphi
,y^{5}=p=t)  \nonumber \\
g_{2} &=&-1,g_{3}=-1,h_{4(0)}=a(u,\lambda ),h_{5(0)}=b(u,\lambda ),%
\mbox{see
(\ref{auxm1})};  \nonumber \\
h_{4} &=&\eta _{4}(u,\lambda ,\varphi )h_{4(0)}(u,\lambda ),h_{5}=\eta
_{5}(u,\lambda ,\varphi )h_{5(0)}(u,\lambda ),  \nonumber \\
\eta _{4} &=&h_{(0)}^{2}\frac{b(u,\lambda )}{a(u,\lambda )}\left\{ \left[
\omega ^{-1}\left( u,\lambda ,\varphi \right) \right] ^{\ast }\right\}
^{2},\eta _{5}=\omega ^{-2}\left( u,\lambda ,\varphi \right) ,
\label{sol4pc} \\
w_{i} &=&0,n_{i}\{x,\omega ,\omega ^{\ast }\}=n_{i}\{u,\lambda ,\omega
\left( u,\lambda ,\varphi \right) ,\omega ^{\ast }\left( u,\lambda ,\varphi
\right) \},\Omega =\Omega _{\lbrack 0]}(u,\lambda )\Omega _{\lbrack
1]}(u,\lambda ,\varphi )  \nonumber \\
\zeta _{i} &=&\left( \partial _{i}\ln |\Omega _{\lbrack 0]}\right) |)~\left(
\ln |\Omega _{\lbrack 1]}|\right) ^{\ast }+\left( \Omega _{\lbrack 1]}^{\ast
}\right) ^{-1}\partial _{i}\Omega _{\lbrack 1]},\eta _{4}a=\Omega _{\lbrack
0]}^{q_{1}/q_{2}}(u,\lambda )\Omega _{\lbrack 1]}^{q_{1}/q_{2}}(u,\lambda
,\varphi ).  \nonumber
\end{eqnarray}
and
\begin{eqnarray}
\mbox{$t_c$--solutions} &:&(x^{2}=u,x^{3}=\lambda ,y^{4}=v=t,y^{5}=p=\varphi
)  \nonumber \\
g_{2} &=&-1,g_{3}=-1,h_{4(0)}=b(u,\lambda ),h_{5(0)}=a(u,\lambda ),%
\mbox{see
(\ref{auxm1})};  \nonumber \\
h_{4} &=&\eta _{4}(u,\lambda ,t)h_{4(0)}(u,\lambda ),h_{5}=\eta
_{5}(u,\lambda ,t)h_{5(0)}(u,\lambda ),  \nonumber \\
\eta _{4} &=&\omega ^{-2}\left( u,\lambda ,t\right) ,\eta _{5}=h_{(0)}^{-2}%
\frac{b(u,\lambda )}{a(u,\lambda )}\left[ \int dt~\omega ^{-1}\left(
u,\lambda ,t\right) \right] ^{2},  \label{sol4tc} \\
w_{i} &=&0,n_{i}\{x,\omega ,\omega ^{\ast }\}=n_{i}\{u,\lambda ,\omega
\left( u,\lambda ,t\right) ,\omega ^{\ast }\left( u,\lambda ,t\right)
\},\Omega =\Omega _{\lbrack 0]}(u,\lambda )\Omega _{\lbrack 1]}(u,\lambda ,t)
\nonumber \\
\zeta _{i} &=&\left( \partial _{i}\ln |\Omega _{\lbrack 0]}\right) |)~\left(
\ln |\Omega _{\lbrack 1]}|\right) ^{\ast }+\left( \Omega _{\lbrack 1]}^{\ast
}\right) ^{-1}\partial _{i}\Omega _{\lbrack 1]},\eta _{4}a=\Omega _{\lbrack
0]}^{q_{1}/q_{2}}(u,\lambda )\Omega _{\lbrack 1]}^{q_{1}/q_{2}}(u,\lambda
,t),  \nonumber
\end{eqnarray}
where the coefficients the $n_{\underline{i}}$ are given by the same
formulas (\ref{nem4}).

Contrary to the solutions (\ref{sol4p1}) and for (\ref{sol4t1}) theirs
conformal anholonomic transforms, respectively, (\ref{sol4pc}) and (\ref
{sol4tc}), can be subjected to such parametrizations of the conformal factor
and conditions on the receptivity $\omega \left( u,\lambda ,v\right) $ as to
obtain in the locally isotropic limit just the Schwarzschild metric (\ref
{schel}). These conditions are stated for $\Omega _{\lbrack
0]}^{q_{1}/q_{2}}=\Omega _{A},$ $\Omega _{\lbrack 1]}^{q_{1}/q_{2}}\eta
_{4}=1,$ $\Omega _{\lbrack 1]}^{q_{1}/q_{2}}\eta _{5}=1,$were $\Omega _{A}$
is that from (\ref{confq}), which is possible if $\eta
_{4}^{-q_{1}/q_{2}}\eta _{5}=1,$which selects a specific form of the
receptivity $\omega .$ \ Putting the values $\eta _{4}$ and $\eta _{5},$
respectively, from (\ref{sol4pc}), or (\ref{sol4tc}), we obtain some
differential, or integral, relations of the unknown $\omega \left(
x^{i},v\right) ,$ which results that
\begin{eqnarray*}
\omega \left( u,\lambda ,\varphi \right) &=&\left( 1-q_{1}/q_{2}\right)
^{-1-q_{1}/q_{2}}\left[ h_{(0)}^{-1}\sqrt{|a/b|}\varphi +\omega _{\lbrack
0]}\left( u,\lambda \right) \right] ,\mbox{ for }\varphi _{c}%
\mbox{--solutions}; \\
\omega \left( u,\lambda ,t\right) &=&\left[ \left( q_{1}/q_{2}-1\right)
h_{(0)}\sqrt{|a/b|}t+\omega _{\lbrack 1]}\left( u,\lambda \right) \right]
^{1-q_{1}/q_{2}},\mbox{ for }t_{c}\mbox{--solutions},
\end{eqnarray*}
for some arbitrary functions $\omega _{\lbrack 0]}\left( u,\lambda \right) $
and $\omega _{\lbrack 1]}\left( u,\lambda \right) .$ The obtained formulas
for $\omega \left( u,\lambda ,\varphi \right) $ and $\omega \left( u,\lambda
,t\right) $ are 4D reductions of the formulas (\ref{cond1a}) and (\ref{cond1}%
).

\subsection{Ellipsiodal 4D metrics of Class B}

We construct another two classes of 4D vacuum solutions which are related to
the metric of Class B (\ref{auxm2}) and which can be reduced to the
Schwarzshild metric in ellipsoidal coordinates (\ref{schel}) by
corresponding parametrizations of receptivity $\omega \left( u,\lambda
,v\right) $. The solutions contain a 2D conformal factor $\varpi $ $%
(u,\lambda )$ for which $\varpi g$ becomes a solution of (\ref{ricci1a}) and
a 4D conformal factor parametrized as $\Omega =\varpi ^{-1}$ $\Omega
_{\lbrack 2]}\left( u,\lambda ,v\right) $ in \ order to set the
constructions into the ansatz (\ref{ansatzc4}) and anholonomic metric
interval (\ref{cdmetric4}).

The data selecting the 4D configurations for $\varphi _{c}$--solutions and $%
t_{c}$--solutions:

\begin{eqnarray}
\mbox{$\varphi_c$--solutions} &:&(x^{2}=u,x^{3}=\lambda ,y^{4}=v=\varphi
,y^{5}=p=t)  \nonumber \\
g_{2} &=&g_{3}=\varpi (u,\lambda )g(u,\lambda ),h_{4(0)}=-\varpi (u,\lambda
),h_{5(0)}=\varpi (u,\lambda )f(u,\lambda ),\mbox{see
(\ref{auxm2})};  \nonumber \\
\varpi &=&g^{-1}\varpi _{0}\exp [a_{2}u+a_{3}\lambda ],~\varpi
_{0},a_{2},a_{3}=const;~\mbox{see
(\ref{solricci1a})}  \nonumber \\
h_{4} &=&\eta _{4}(u,\lambda ,\varphi )h_{4(0)}(u,\lambda ),h_{5}=\eta
_{5}(u,\lambda ,\varphi )h_{5(0)}(u,\lambda ),  \nonumber \\
\eta _{4} &=&-h_{(0)}^{2}f(u,\lambda )\left\{ \left[ \omega ^{-1}\left(
u,\lambda ,\varphi \right) \right] ^{\ast }\right\} ^{2},\eta _{5}=\omega
^{-2}\left( u,\lambda ,\varphi \right) ,  \label{sol4p} \\
w_{i} &=&0,n_{i}\{x,\omega ,\omega ^{\ast }\}=n_{i}\{u,\lambda ,\omega
\left( u,\lambda ,\varphi \right) ,\omega ^{\ast }\left( u,\lambda ,\varphi
\right) \},\Omega =\varpi ^{-1}(u,\lambda )\Omega _{\lbrack 2]}(u,\lambda
,\varphi )  \nonumber \\
\zeta _{\underline{i}} &=&\partial _{\underline{i}}\ln |\varpi |)~\left( \ln
|\Omega _{\lbrack 2]}|\right) ^{\ast }+\left( \Omega _{\lbrack 2]}^{\ast
}\right) ^{-1}\partial _{\underline{i}}\Omega _{\lbrack 2]},\eta
_{4}=-\varpi ^{-q_{1}/q_{2}}(u,\lambda )\Omega _{\lbrack
2]}^{q_{1}/q_{2}}(u,\lambda ,\varphi ).  \nonumber
\end{eqnarray}
and
\begin{eqnarray}
\mbox{$t_c$--solutions} &:&(x^{2}=u,x^{3}=\lambda ,y^{4}=v=t,y^{5}=p=\varphi
)  \nonumber \\
g_{2} &=&g_{3}=\varpi (u,\lambda )g(u,\lambda ),h_{4(0)}=\varpi (u,\lambda
)f(u,\lambda ),h_{5(0)}=-\varpi (u,\lambda ),\mbox{see
(\ref{auxm2})};  \nonumber \\
\varpi &=&g^{-1}\varpi _{0}\exp [a_{2}u+a_{3}\lambda ],~\varpi
_{0},a_{2},a_{3}=const,~\mbox{see
(\ref{solricci1a})}  \nonumber \\
h_{4} &=&\eta _{4}(u,\lambda ,t)h_{4(0)}(x^{i}),h_{5}=\eta _{5}(u,\lambda
,t)h_{5(0)}(x^{i}),  \nonumber \\
\eta _{4} &=&\omega ^{-2}\left( u,\lambda ,t\right) ,\eta
_{5}=-h_{(0)}^{-2}f(u,\lambda )\left[ \int dt~\omega ^{-1}\left( u,\lambda
,t\right) \right] ^{2},  \label{sol4t} \\
w_{i} &=&0,n_{i}\{x,\omega ,\omega ^{\ast }\}=n_{i}\{u,\lambda ,\omega
\left( u,\lambda ,t\right) ,\omega ^{\ast }\left( u,\lambda ,t\right)
\},\Omega =\varpi ^{-1}(u,\lambda )\Omega _{\lbrack 2]}(u,\lambda ,t)
\nonumber \\
\zeta _{i} &=&\partial _{i}(\ln |\varpi |)~\left( \ln |\Omega _{\lbrack
2]}|\right) ^{\ast }+\left( \Omega _{\lbrack 2]}^{\ast }\right)
^{-1}\partial _{i}\Omega _{\lbrack 2]},\eta _{4}=-\varpi
^{-q_{1}/q_{2}}(u,\lambda )\Omega _{\lbrack 2]}^{q_{1}/q_{2}}(u,\lambda ,t).
\nonumber
\end{eqnarray}
where the coefficients $n_{i}$ can be found explicitly by introducing the
corresponding values $\eta _{4}$ and $\eta _{5}$ in formula (\ref{nel}).

For the 4D Class B solutions one can be imposed some conditions (see
previous subsection) when the solutions (\ref{sol4p}) and (\ref{sol4t}) have
in the locally anisotropic limit the Schwarzshild solution, which imposes
some specific parametrizations and dependencies on $\Omega _{\lbrack
2]}(u,\lambda ,v)$ and $\omega \left( u,\lambda ,v\right) $ like (\ref
{cond1a}) and (\ref{cond1}). We omit these considerations because for
aholonomic configurations and nonlinear solutions there are not arguments to
prefer any holonomic limits of such off--diagonal metrics.

We conclude this Section by noting that for the considered classes of
ellipsoidal black hole 4D solutions the so--called $t$--component of metric
contains modifications of the Schwarzschild potential
\[
\Phi =-\frac{M}{M_{P[4]}^{2}r}\mbox{ into }\Phi =-\frac{M\omega \left(
u,\lambda ,v\right) }{M_{P[4]}^{2}r},
\]
where $M_{P[4]}$ is the usual 4D Plank constant; the metric coefficients are
given with respect to the corresponding aholonomic frame of reference. In 4D
anholonomic gravity the receptivity $\omega \left( u,\lambda ,v\right) $ is
considered to renormalize the mass constant. Such gravitational
self-polarizations are induced by anholonomic vacuum gravitational
interactions. They should be defined experimentally or computed following a
model of quantum gravity.

\section{The Cosmological Constant and Anisotropy}

In this Section we analyze the general properties of anholonomic
Einstein equations in 5D and 4D gravity with cosmological
constant and construct a 5D exact solution with cosmological
constant.

\subsection{4D and 5D Anholnomic Einstein spaces}

There is a difference between locally anisotropic 4D and 5D
gravity. The first theory admits an ''isotropic'' 4D cosmological
constant $\Lambda _{\lbrack 4]}=\Lambda $ even for anisotropic
gravitational configurations. The second, 5D, theory admits
extensions of vacuum anistoropic solutions to those with a
cosmological constant only for anisotropic 5D sources
parametrized like $\Lambda _{\lbrack 5]\alpha \beta }=(2\Lambda
g_{11},\Lambda g_{\underline{\alpha }\underline{\beta }})$ (see
the Corollary 4 below). We emphasize that the conclusions from
this subsection refer to the two classes of ansatz (\ref{ansatz})
and (\ref{ansatzc}).

The simplest way to consider a source into the 4D Einstein
equations, both with or not anistoropy, is to consider a
gravitational constant $\Lambda $ and to write the field equations
\begin{equation}
G_{\underline{\beta }}^{\underline{\alpha }}=\Lambda _{\lbrack 4]}\delta _{%
\underline{\beta }}^{\underline{\alpha }}  \label{einst4cc}
\end{equation}
which means that we introduced a ''vacuum'' energy--momentum tensor $\kappa
\Upsilon _{\underline{\beta }}^{\underline{\alpha }}=\Lambda _{\lbrack
4]}\delta _{\underline{\beta }}^{\underline{\alpha }}$ which is diagonal
with respect to anholonomic frames and the conditions (\ref{emcond4})
transforms into $\Upsilon _{2}^{2}=\Upsilon _{3}^{3}=\Upsilon
_{4}^{4}=\Upsilon _{5}^{5}=\kappa ^{-1}\Lambda .$ According to A. Z. Petrov
\cite{petrov} the spaces described by solutions of the Einstein equations
\[
R_{\underline{\alpha }\underline{\beta }}=\Lambda g_{\underline{\alpha }%
\underline{\beta }},\Lambda =const
\]
are called the Einstein spaces. With respect to anisotropic frames we shall
use the term anholonomic (equivalently, anisotropic) Einstein spaces.

In order to extend the equations (\ref{einst4cc}) to 5D gravity we have to
take into consideration the compatibility conditions for the
energy--momentum tensors (\ref{emcond}).

\begin{corollary}
We are able to satisfy the conditions of the Corollary 2 if \ we consider a
5D diagonal source $\Upsilon _{\beta }^{\alpha }=\{2\Lambda ,$ $\Upsilon _{%
\underline{\beta }}^{\underline{\alpha }}=\Lambda \delta _{\underline{\beta }%
}^{\underline{\alpha }}\},$ for an anisotropic 5D cosmological constant source $%
(2\Lambda g_{11},\Lambda g_{\underline{\alpha }\underline{\beta
}}).$ The 5D Einstein equations with anisotropic cosmological
"constants", for ansatz (\ref {ansatz}) are written in the form
\begin{equation}
R_{2}^{2}=S_{4}^{4}=-\Lambda .  \label{eecvcosm}
\end{equation}
These equations without coordinate $x^{1}$ and $g_{11}$ hold for the (\ref
{ansatz4}). We can extend the constructions for the ansatz with conformal
factors, (\ref{ansatzc}) and (\ref{ansatzc4}) by considering additional
coefficients $\zeta _{i}$ satisfying the equations (\ref{confeq}) and (\ref
{confeq4}) for non vanishing values of $w_{i}.$
\end{corollary}

The proof follows from Corollaries 1 and 2 formulated respectively to 4D and
5D gravity (see formulas (\ref{einstdiag4}) and (\ref{emcond4}) and,
correspondingly, (\ref{einstdiag}) and (\ref{emcond})).

\begin{theorem}
The nontrivial components of the 5D Einstein equations with anisotropic
cosmological constant, $R_{11}=2\Lambda g_{11}$ and $R_{\underline{\alpha }%
\underline{\beta }}=\Lambda g_{\underline{\alpha }\underline{\beta }},$ for
the ansatz (\ref{ansatzc}) and anholonomic metric (\ref{cdmetric}) given
with respect to anholonomic frames (\ref{ddif2}) and (\ref{dder2}) are
written in a form with separation of variables:
\begin{eqnarray}
g_{3}^{\bullet \bullet }-\frac{g_{2}^{\bullet }g_{3}^{\bullet }}{2g_{2}}-%
\frac{(g_{3}^{\bullet })^{2}}{2g_{3}}+g_{2}^{^{\prime \prime }}-\frac{%
g_{2}^{^{\prime }}g_{3}^{^{\prime }}}{2g_{3}}-\frac{(g_{2}^{^{\prime }})^{2}%
}{2g_{2}} &=&2\Lambda g_{2}g_{3},  \label{ricci1const} \\
h_{5}^{\ast \ast }-h_{5}^{\ast }[\ln \sqrt{|h_{4}h_{5}|}]^{\ast }
&=&2\Lambda h_{4}h_{5},  \label{ricci2const} \\
w_{i}\beta +\alpha _{i} &=&0,  \label{ricci3const} \\
n_{i}^{\ast \ast }+\gamma n_{i}^{\ast } &=&0,  \label{ricci4const} \\
\partial _{i}\Omega -(w_{i}+\zeta _{{i}})\Omega ^{\ast } &=&0.
\label{confeql}
\end{eqnarray}
where
\begin{equation}
\alpha _{i}=\partial _{i}{h_{5}^{\ast }}-h_{5}^{\ast }\partial _{i}\ln \sqrt{%
|h_{4}h_{5}|},\beta =2\Lambda h_{4}h_{5},\gamma =3h_{5}^{\ast
}/2h_{5}-h_{4}^{\ast }/h_{4}.  \label{abcl4}
\end{equation}
\end{theorem}

The Theorem 4 is a generalization of the Theorem 2 for energy--momentum
tensors induced by the an anisotropic 5D constant. The proof follows from (%
\ref{ricci1a})--(\ref{ricci4a}) and (\ref{confeq}), revised as to
satisfy the formulas\ (\ref{ricci1b}) and (\ref{ricci2b}) with
that substantial difference that $\beta \neq 0$ and in this case,
in general, $w_{i}\neq 0.$ We conclude that in the presence of a
nonvanishing cosmological constant the
equations (\ref{ricci1a}) and (\ref{ricci2a}) transform respectively into (%
\ref{ricci1const}) and \ (\ref{ricci2const}) \ which have a more
general nonlinearity because of  the $2\Lambda g_{2}g_{3}$ and
$2\Lambda
h_{4}h_{5}$ terms. For instance, the solutions with $g_{2}=const$ and $%
g_{3}=const$ (and $h_{4}=const$ and $h_{5}=const)$ are not admitted. This
makes more sophisticate the procedure of definition of $g_{2}$ for a given $%
g_{3}$ (or inversely, of definition of $g_{3}$ for a given $g_{2})$ from (%
\ref{ricci1const}) [similarly of construction $h_{4}$ for a given
$h_{5}$ from (\ref{ricci2const}) and inversely], nevertheless,
the separation of variables is not affected by introduction of
cosmological constant and there is a number of possibilities to
generate new exact solutions.

The general properties of solutions of the system (\ref{ricci1const})--(\ref
{confeql}) are stated by the

\begin{theorem}
The system of second order nonlinear partial differential equations (\ref
{ricci1const})-(\ref{ricci4const}) and (\ref{confeql}) can be solved in
general form if there are given some values of functions $g_{2}(x^{2},x^{3})$
(or $g_{3}(x^{2},x^{3})),h_{4}\left( x^{i},v\right) $ (or $h_{5}\left(
x^{i},v\right) )$ and $\Omega \left( x^{i},v\right) :$

\begin{itemize}
\item  The general solution of equation (\ref{ricci1const}) is to be found
from the equation
\begin{equation}
\varpi \varpi ^{\bullet \bullet }-(\varpi ^{\bullet })^{2}+\varpi \varpi
^{^{\prime \prime }}-(\varpi ^{^{\prime }})^{2}=2\Lambda \varpi ^{3}.
\label{auxr1}
\end{equation}
for a coordinate transform coordinate transforms $x^{2,3}\rightarrow
\widetilde{x}^{2,3}\left( u,\lambda \right) $ for which
\[
g_{2}(u,\lambda )(du)^{2}+g_{3}(u,\lambda )(d\lambda )^{2}\rightarrow \varpi
\left[ (d\widetilde{x}^{2})^{2}+\epsilon (d\widetilde{x}^{3})^{2}\right]
,\epsilon =\pm 1
\]
and $\varpi ^{\bullet }=\partial \varpi /\partial \widetilde{x}^{2}$ and $%
\varpi ^{^{\prime }}=\partial \varpi /\partial \widetilde{x}^{3}.$

\item  The equation (\ref{ricci2const}) relates two functions $h_{4}\left(
x^{i},v\right) $ and $h_{5}\left( x^{i},v\right) $ with $h_{5}^{\ast }\neq
0. $ If the function $h_{5}$ is given we can find $h_{4}$ as a solution of
\begin{equation}
h_{4}^{\ast }+\frac{2\Lambda }{\tau }(h_{4})^{2}+2\left( \frac{\tau ^{\ast }%
}{\tau }-\tau \right) h_{4}=0,  \label{auxr2c}
\end{equation}

where $\tau =h_{5}^{\ast }/2h_{5}.$

\item  The exact solutions of (\ref{ricci3const}) for $\beta \neq 0$ is
\begin{eqnarray}
w_{k} &=&-\alpha _{k}/\beta ,  \label{aw} \\
&=&\partial _{k}\ln [\sqrt{|h_{4}h_{5}|}/|h_{5}^{\ast }|]/\partial _{v}\ln [%
\sqrt{|h_{4}h_{5}|}/|h_{5}^{\ast }|],  \nonumber
\end{eqnarray}
for $\partial _{v}=\partial /\partial v$ and $h_{5}^{\ast }\neq 0.$

\item  The exact solution of (\ref{ricci4const}) is
\begin{eqnarray}
n_{k} &=&n_{k[1]}\left( x^{i}\right) +n_{k[2]}\left( x^{i}\right) \int
[h_{4}/(\sqrt{|h_{5}|})^{3}]dv,  \label{nlambda} \\
&=&n_{k[1]}\left( x^{i}\right) +n_{k[2]}\left( x^{i}\right) \int [1/(\sqrt{%
|h_{5}|})^{3}]dv,~h_{4}^{\ast }=0,  \nonumber
\end{eqnarray}
for some functions $n_{k[1,2]}\left( x^{i}\right) $ stated by boundary
conditions.

\item  The exact solution of (\ref{confeq}) is given by
\begin{equation}
\zeta _{i}=-w_{i}+(\Omega ^{\ast })^{-1}\partial _{i}\Omega ,\quad \Omega
^{\ast }\neq 0,  \label{aconf4}
\end{equation}
\end{itemize}
\end{theorem}

We note that by a corresponding re--parametrizations of the conformal factor
$\Omega \left( x^{i},v\right) $ we can reduce (\ref{auxr1}) to
\begin{equation}
\varpi \varpi ^{\bullet \bullet }-(\varpi ^{\bullet })^{2}=2\Lambda \varpi
^{3}  \label{redaux}
\end{equation}
which has an exact solution $\varpi =\varpi \left( \widetilde{x}%
^{2}\right) $ to be found from
\[
(\varpi ^{\bullet })^{2}=\varpi ^{3}\left( C\varpi ^{-1}+4\Lambda \right)
,C=const,
\]
(or, inversely, to reduce to
\[
\varpi \varpi ^{^{\prime \prime }}-(\varpi ^{^{\prime }})^{2}=2\Lambda
\varpi ^{3}
\]
with exact solution $\varpi =\varpi \left( \widetilde{x}^{3}\right) $ found
from
\[
(\varpi ^{\prime })^{2}=\varpi ^{3}\left( C\varpi ^{-1}+4\Lambda \right)
,C=const).
\]
The inverse problem of definition of $h_{5}$ for a given $h_{4}$ can be
solved in explicit form when $h_{4}^{\ast }=0,$ $h_{4}=h_{4(0)}(x^{i}).$ In
this case we have to solve
\begin{equation}
h_{5}^{\ast \ast }+\frac{(h_{5}^{\ast })^{2}}{2h_{5}}-2\Lambda
h_{4(0)}h_{5}=0,  \label{auxr2ccp}
\end{equation}
which admits exact solutions by reduction to a Bernulli equation.

The proof of Theorem 5 is outlined in Appendix C.

The conditions of the Theorem 4 and 5 can be reduced to 4D anholonomic
spacetimes with ''isotropic'' cosmological constant $\Lambda .$ To do this
we have to eliminate dependencies on the coordinate $x^{1}$ and to consider
the 4D ansatz without $g_{11}$ term as it was stated in the subsection II E.

\subsection{5D anisotropic black holes with cosmological constant}

We give an example of generalization of ansiotropic black hole
solutions of Class A , constructed in the Section III, as they
will contain the
cosmological constant $\Lambda ;$ we extend the solutions given by the data (%
\ref{sol5pc}).

Our new 5D $\varphi $-- solution is parametrized by an ansatz with conformal
factor $\Omega (x^{i},v)$ (see (\ref{ansatzc}) and (\ref{cdmetric})) as $%
\Omega =\varpi ^{-1}(u)\Omega _{\lbrack 0]}(x^{i})\varpi
^{-1}(u)\Omega _{\lbrack 1]}(x^{i},v).$ The factor $\varpi (u)$
is chosen  to be a
solution of (\ref{redaux}). This conformal data must satisfy the conditions (%
\ref{confq}) and (\ref{confsol}), i. e.
\[
\varpi ^{-q_{1}/q_{2}}\Omega _{\lbrack 0]}^{q_{1}/q_{2}}\Omega _{\lbrack
1]}^{q_{1}/q_{2}}=\eta _{4}\varpi h_{4(0)}
\]
for some integers $q_{1}$ and $q_{2},$ where $\eta _{4}$ is found as $%
h_{4}=\eta _{4}\varpi h_{4(0)}$ is a solution of equation (\ref{auxr2c}). The factor
 $\Omega _{\lbrack 0]}(x^{i})$ could be chosen as to obtain in the locally
isotropic limit and $\Lambda \rightarrow 0$ the Scwarzshild
metric in ellipsoidal coordinates (\ref{schel}). Putting
$h_{5}=\eta _{5}\varpi h_{5(0)},$ $\eta _{5}h_{5(0)}$ in the
ansatz for (\ref{sol5pc}), for which we compute the value $\tau
=h_{5}^{\ast }/2h_{5},$ we obtain from (\ref
{auxr2c}) an equation for $\eta _{4},$%
\[
\eta _{4}^{\ast }+\frac{2\Lambda }{\tau }\varpi h_{4(0)}(\eta
_{4})^{2}+2\left( \frac{\tau ^{\ast }}{\tau }-\tau \right) \eta _{4}=0
\]
which is a Bernulli equation \cite{kamke} and admit an exact solution, in
general, in non explicit form, $\eta _{4}=\eta _{4}^{[bern]}(x^{i},v,\Lambda
,\varpi ,\omega ,a,b),$ were we emphasize the functional dependencies on
functions $\varpi ,\omega ,a,b$ and cosmological constant $\Lambda .$ Having
defined $\eta _{4[bern]},$ $\eta _{5}$ and $\varpi ,$ we can compute the $%
\alpha _{i}$--$,\beta -,$ and $\gamma $--coefficients, expressed as $\alpha
_{i}=\alpha _{i}^{[bern]}(x^{i},v,\Lambda ,\varpi ,\omega ,a,b),\beta =\beta
^{\lbrack bern]}(x^{i},v,\Lambda ,\varpi ,\omega ,a,b)$ and $\gamma =\gamma
^{\lbrack bern]}(x^{i},v,\Lambda ,\varpi ,\omega ,a,b),$ following the
formulas (\ref{abcl4}).

The next step is to find
\[
w_{i}=w_{i}^{[bern]}(x^{i},v,\Lambda ,\varpi ,\omega ,a,b)\mbox{ and }%
n_{i}=n_{i}^{[bern]}(x^{i},v,\Lambda ,\varpi ,\omega ,a,b)
\]
as for the general solutions (\ref{aw}) and (\ref{nlambda}).

At the final step we are able to compute the the second anisotropy
coefficients
\[
\zeta _{i}=-w_{i}^{[bern]}+\left( \partial _{i}\ln |\varpi ^{-1}\Omega
_{\lbrack 0]}\right) |)~\left( \ln |\Omega _{\lbrack 1]}|\right) ^{\ast
}+\left( \Omega _{\lbrack 1]}^{\ast }\right) ^{-1}\partial _{i}\Omega
_{\lbrack 1]},
\]
which depends on an arbitrary function $\Omega _{\lbrack 0]}(u,\lambda ).$
If we state $\Omega _{\lbrack 0]}(u,\lambda )=\Omega _{A},$ as for $\Omega
_{A}$ from (\ref{auxm2}), see similar details with respect to formulas (\ref
{cond1a}), (\ref{cond2}) and (\ref{cond1}).

The data for the exact solutions with cosmological constant for $v=\varphi $
can be stated in the form
\begin{eqnarray}
\mbox{$\varphi_c$--solutions} &:&(x^{1}=\chi ,x^{2}=u,x^{3}=\lambda
,y^{4}=v=\varphi ,y^{5}=p=t),g_{1}=\pm 1,  \nonumber \\
g_{2} &=&\varpi (u),g_{3}=\varpi (u),h_{4(0)}=a(u,\lambda
),h_{5(0)}=b(u,\lambda ),\mbox{see (\ref{auxm1}) and  (\ref{redaux})};
\nonumber \\
h_{4} &=&\eta _{4}(x^{i},\varphi )\varpi (u)h_{4(0)}(x^{i}),h_{5}=\eta
_{5}(x^{i},\varphi )\varpi (u)h_{5(0)}(x^{i}),  \nonumber \\
\eta _{4} &=&\eta _{4}^{[bern]}(x^{i},v,\Lambda ,\varpi ,\omega ,a,b),\eta
_{5}=\omega ^{-2}\left( \chi ,u,\lambda ,\varphi \right) ,  \label{slambdap1}
\\
w_{i} &=&w_{i}^{[bern]}(x^{i},v,\Lambda ,\varpi ,\omega
,a,b),n_{i}\{x,\omega ,\omega ^{\ast }\}=n_{i}^{[bern]}(x^{i},v,\Lambda
,\varpi ,\omega ,a,b),  \nonumber \\
\Omega &=&\varpi ^{-1}(u)\Omega _{\lbrack 0]}(x^{i})\Omega _{\lbrack
1]}(x^{i},\varphi ),\eta _{4}a=\Omega _{\lbrack
0]}^{q_{1}/q_{2}}(x^{i})\Omega _{\lbrack 1]}^{q_{1}/q_{2}}(x^{i},\varphi ).
\nonumber \\
\zeta _{i} &=&-w_{i}^{[bern]}+\left( \partial _{i}\ln |\varpi ^{-1}\Omega
_{\lbrack 0]}\right) |)~\left( \ln |\Omega _{\lbrack 1]}|\right) ^{\ast
}+\left( \Omega _{\lbrack 1]}^{\ast }\right) ^{-1}\partial _{i}\Omega
_{\lbrack 1]}.  \nonumber
\end{eqnarray}

We note that a solution with $v=t$ can be constructed as to
generalize (\ref {sol5tc}) in order to contain $\Lambda .$ We can
not present such data in explicit form because in this case we
have to define $\eta _{5}$ by integrating an equation like
(\ref{ricci2const}) for $h_{5},$ for a given $h_{4},$ with
$h_{4}^*\neq 0$ which can not be integrated in explicit form.

The solution (\ref{slambdap1}) has has the same the two very
interesting properties as the solution (\ref{sol5pc}):\
 1) it admits a warped factor on the 5th coordinate, like $%
\Omega _{\lbrack 1]}^{q_{1}/q_{2}}\sim \exp [-k|\chi |],$ which
in this case is constructed for an anisotropic 5D vacuum
gravitational configuration with anisotropic cosmological
constant and does not follow from a brane configuration like in
Refs. \cite{rs}; 2) we can impose such conditions on the
receptivity $\omega \left( x^{i},\varphi \right) $ as to obtain
in the locally isotropic limit just the Schwarzschild metric
(\ref{schel}) trivially embedded into the 5D spacetime (the
procedure is the same as in the subsection IIIB).

Finally, we note that in a similar manner like in the Sections III and IV we
can construct another classes of anisotropic black holes solutions in 5D and
4D spacetimes with cosmological constants, being of Class A or Class B, with
anisotropic $\varphi $--coordinate, or anisotropic $t$--coordinate. We omit
the explicit data which are some nonlinear anholonomic generalizations of
those solutions.

\section{Conclusions}

We formulated a new method of constructing exact solutions of
Einstein equations with off--diagonal metrics in 4D and 5D
gravity. We introduced ahnolonomic transforms which diagonalize
metrics and simplify  the system of  gravitational field
equations.   The method works also for gravitational
configurations with cosmological constants and for non--trivial
matter sources. We constructed different classes of new exact
solutions of the Einstein equations is 5D and 4D gravity which
describe a generic anholonomic (anisotropic) dynamics modeled by
off--diagonal metrics and anholonomic frames with mixed holonomic
and anholonomic variables. They extend the class of exact
solutions with linear extensions to the bulk 5D gravity
\cite{gian}.

We emphasized such exact solutions which can be associated to
some black hole like configurations in 5D and 4D gravity. We
consider that the constructed off--diagonal metrics define
anisotropic black holes because they have a static horizon
parametrized by a rotation ellipsoid hypersurface, they are
singular in focuses of ellipsoid (or on the circle of focuses,
for flattened ellipsoids) and they reduce in the locally
anisotropic limit, with holonomic coordinates, to the
Schwarzshild solution in ellipsoidal coordinates, or to some
conformal transforms of the  Schwarzshild metric.

The new classes of solutions admit variations of constants (in
time and extra dimension coordinate) and anholonomic
gravitational polarizations of masses which are induced by
nonlinear gravitational interactions in the bulk of 5D gravity
and by a constrained (anholonomic) dynamics of the fields in the
4D gravity. There are possible solutions with warped factors
which are defined by some\ vacuum 5D gravitational interactions
in the bulk and not by a specific brane configuration with
energy--momentum tensor source. \ We emphasized anisotropies
which in the effective 4D spacetime preserve the local Lorentz
invariance but the method allows constructions with violation of
local Lorentz symmetry like in Refs.  \cite{csaki}. In order to
generate such solutions we should admit that the metric
coefficients depends, for instance, anisotropically on extra
dimension coordinate. \

It should be noted that the anholonomic frame method deals with
generic off--diagonal metrics and nonlinear systems of equations
and allows to construct substantially nonlinear solutions. In
general, such solutions could not have a locally isotropic limit
with a holonomic analog. We can understand the physical
properties of such solutions by analyzing both the metric
coefficients stated with respect to an adapted anholonomic frame
of reference and by a study of the coefficients defining such
frames.

There is a subclass of static anisotropic black holes solutions,
with static ellipsoidal horizons, which do not violate the well
known Israel and Carter theorems \cite{israel} on spherical
symmetry of solutions in assymptotically flat spacetimes. Those
theorems were proved in the radial symmetry asymptotic limit and
for holonomic coordinates. There is not a much difference between
\ 3D static spherical and ellipsoidal horizons at long distances.
In other turn, the statements of the mentioned theorems do not
refers to generic off--diagonal gravitational metrics,
anholonomic frames and anholonomic deformations of symmetries.

Finally, we note that the anholonomic frame method may have a
number of applications in modern brane and string/M--theory
gravity because it defines a general formalism of constructing
exact solutions with off--diagonal metrics. It results in such
prescriptions on anholonomic ''mappings'' of some known locally
isotropic solutions from a gravity/string theory that new types of
anisotropic solutions are generated:

{\em A vacuum, or non-vacuum, solution, and metrics conformally
equivalent to a such solution, parametrized by a diagonal matrix
given with respect to a holonomic (coordinate) base, contained in
a trivial form of ansatz (\ref {ansatz}), or (\ref{ansatzc}), can
be generalized to an anisotropic solution with similar but
anisotropically renormalized physical constants and diagonal
metric coefficients given with respect to adapted anholonomic
frames; the new anholonomic metric defines an exact solution of a
simplified form of \ the Einstein equations
(\ref{ricci1a})--(\ref{ricci4a}) and (\ref {confeq}); such
solutions are parametrized by off--diagonal metrics if they are
re--defined with respect to coordinate frames }.

\subsection*{Acknowledgements}

The author thanks D. Singleton, E. Gaburov, D. Gontsa and Nadejda
Vicol for collaboration and discussing of results. The work is
supported both by a 2000--2001 California State University
Legislative Award and a NATO/Portugal fellowship grant at the
Technical University of Lisbon.

\appendix

\section{ Anholonomic Frames and Nonlinear Connections}

For convenience, we outline here the basic formulas for connections,
curvatures and, induced by anholonomic frames, torsions on (pseudo)
Riemannian spacetimes provided with N--coefficient bases (\ref{dder}) and (%
\ref{ddif}) \cite{v,v1}. The N--coefficients define an associated nonlinear
connection (in brief, N--connection) structure. On (pseudo)--Riemannian
spacetimes the  N--connection structure can be treated as a ''pure''
anholonomic frame effect which is induced if we are dealing with mixed sets
of holonomic--anholonomic basis vectors. When we are transferring our
considerations only to coordinate frames (\ref{pder}) and (\ref{pdif}) the
N--connection coefficients are removed into both off--diagonal and diagonal
components of the metric like in (\ref{ansatz}). In some cases the
N--connection (anholonomic) structure is to be stated in a non--dynamical
form by definition of some initial (boundary) conditions for the frame
structure, following some prescribed symmetries of the gravitational--matter
field interactions, or , in another cases, a subset of N--coefficients have
to be treated as some dynamical variables defined as to satisfy the Einstein
equations.

\subsection{D--connections, d--torsions and d--curvatures}

If a pseudo--Riemannian spacetime is enabled with a N--connection strucutre,
the components of geometrical objects (for instance, linear connections and
tensors) are distinguished into horizontal components (in brief
h--components, labeled by indices like $i,j,k,...)$ and vertical components
(in brief v--components, labeled by indices like $a,b,c,..).$ One call such
objects, distinguished (d) by the N--connection structure, as d--tensors,
d--connections, d--spinors and so on \cite{ma,vf,v}.

\subsubsection{D--metrics and d-connections:}

A metric of type (\ref{dmetric}), in general, with arbitrary coefficients $%
g_{ij}\left( x^k,y^a\right) $ and $h_{ab}\left( x^k,y^a\right) $ defined
with respect to a N--elongated basis (\ref{ddif}) is called a d--metric.

A linear connection $D_{\delta _{\gamma }}\delta _{\beta }=\Gamma _{\ \beta
\gamma }^{\alpha }\left( x,y\right) \delta _{\alpha },$ associated to an
operator of covariant derivation $D,$ is compatible with a metric $g_{\alpha
\beta }$ and N--connection structure on a 5D pseudo--Riemannian spacetime if
$D_{\alpha }g_{\beta \gamma }=0.$ The linear d--connection is parametrized
by irreducible h--v--components,\ $\Gamma _{\ \beta \gamma }^{\alpha
}=\left( L_{\ jk}^{i},L_{\ bk}^{a},C_{\ jc}^{i},C_{\ bc}^{a}\right) ,$ where
\begin{eqnarray}
L_{\ jk}^{i} &=&\frac{1}{2}g^{in}\left( \delta _{k}g_{nj}+\delta
_{j}g_{nk}-\delta _{n}g_{jk}\right) ,  \label{dcon} \\
L_{\ bk}^{a} &=&\partial _{b}N_{k}^{a}+\frac{1}{2}h^{ac}\left( \delta
_{k}h_{bc}-h_{dc}\partial _{b}N_{k}^{d}-h_{db}\partial _{c}N_{k}^{d}\right) ,
\nonumber \\
C_{\ jc}^{i} &=&\frac{1}{2}g^{ik}\partial _{c}g_{jk},\ C_{\ bc}^{a}=\frac{1}{%
2}h^{ad}\left( \partial _{c}h_{db}+\partial _{b}h_{dc}-\partial
_{d}h_{bc}\right) .  \nonumber
\end{eqnarray}
This defines a canonical linear connection (distinguished by a
N--connection, in brief, the canonical d--connection) which is similar to
the metric connection introduced by Christoffel symbols in the case of
holonomic bases.

\subsubsection{D--torsions and d--curvatures:}

The anholonomic coefficients $W_{\ \alpha \beta }^{\gamma }$ and
N--elongated derivatives give nontrivial coefficients for the torsion
tensor, $T(\delta _{\gamma },\delta _{\beta })=T_{\ \beta \gamma }^{\alpha
}\delta _{\alpha },$ where
\begin{equation}
T_{\ \beta \gamma }^{\alpha }=\Gamma _{\ \beta \gamma }^{\alpha }-\Gamma _{\
\gamma \beta }^{\alpha }+w_{\ \beta \gamma }^{\alpha },  \label{torsion}
\end{equation}
and for the curvature tensor, $R(\delta _{\tau },\delta _{\gamma })\delta
_{\beta }=R_{\beta \ \gamma \tau }^{\ \alpha }\delta _{\alpha },$ where
\begin{eqnarray}
R_{\beta \ \gamma \tau }^{\ \alpha } &=&\delta _{\tau }\Gamma _{\ \beta
\gamma }^{\alpha }-\delta _{\gamma }\Gamma _{\ \beta \tau }^{\alpha }
\nonumber \\
&&+\Gamma _{\ \beta \gamma }^{\varphi }\Gamma _{\ \varphi \tau }^{\alpha
}-\Gamma _{\ \beta \tau }^{\varphi }\Gamma _{\ \varphi \gamma }^{\alpha
}+\Gamma _{\ \beta \varphi }^{\alpha }w_{\ \gamma \tau }^{\varphi }.
\label{curvature}
\end{eqnarray}
We emphasize that the torsion tensor on (pseudo) Riemannian spacetimes is
induced by anholonomic frames, whereas its components vanish with respect to
holonomic frames. All tensors are distinguished (d) by the N--connection
structure into irreducible h--v--components, and are called d--tensors. For
instance, the torsion, d--tensor has the following irreducible,
nonvanishing, h--v--components,\ $T_{\ \beta \gamma }^{\alpha }=\{T_{\
jk}^{i},C_{\ ja}^{i},S_{\ bc}^{a},T_{\ ij}^{a},T_{\ bi}^{a}\},$ where
\begin{eqnarray}
T_{.jk}^{i} &=&T_{jk}^{i}=L_{jk}^{i}-L_{kj}^{i},\quad
T_{ja}^{i}=C_{.ja}^{i},\quad T_{aj}^{i}=-C_{ja}^{i},  \nonumber \\
T_{.ja}^{i} &=&0,\quad T_{.bc}^{a}=S_{.bc}^{a}=C_{bc}^{a}-C_{cb}^{a},
\label{dtors} \\
T_{.ij}^{a} &=&-\Omega _{ij}^{a},\quad T_{.bi}^{a}=\partial
_{b}N_{i}^{a}-L_{.bi}^{a},\quad T_{.ib}^{a}=-T_{.bi}^{a}  \nonumber
\end{eqnarray}
(the d--torsion is computed by substituting the h--v--components of the
canonical d--connection (\ref{dcon}) and anholonomic coefficients (\ref
{anholonomy}) into the formula for the torsion coefficients (\ref{torsion}%
)), where
\[
\Omega _{ij}^{a}=\delta _{j}N_{i}^{a}-\delta _{i}N_{j}^{a}
\]
is called the N--connection curvature (N--curvature).

The curvature d-tensor has the following irreducible, non-vanishing,
h--v--components\ $R_{\beta \ \gamma \tau }^{\ \alpha
}=%
\{R_{h.jk}^{.i},R_{b.jk}^{.a},P_{j.ka}^{.i},P_{b.ka}^{.c},S_{j.bc}^{.i},S_{b.cd}^{.a}\},
$\ where
\begin{eqnarray}
R_{h.jk}^{.i} &=&\delta _{k}L_{.hj}^{i}-\delta
_{j}L_{.hk}^{i}+L_{.hj}^{m}L_{mk}^{i}-L_{.hk}^{m}L_{mj}^{i}-C_{.ha}^{i}%
\Omega _{.jk}^{a},  \label{dcurvatures} \\
R_{b.jk}^{.a} &=&\delta _{k}L_{.bj}^{a}-\delta
_{j}L_{.bk}^{a}+L_{.bj}^{c}L_{.ck}^{a}-L_{.bk}^{c}L_{.cj}^{a}-C_{.bc}^{a}%
\Omega _{.jk}^{c},  \nonumber \\
P_{j.ka}^{.i} &=&\partial _{a}L_{.jk}^{i}+C_{.jb}^{i}T_{.ka}^{b}-(\delta
_{k}C_{.ja}^{i}+L_{.lk}^{i}C_{.ja}^{l}-L_{.jk}^{l}C_{.la}^{i}-L_{.ak}^{c}C_{.jc}^{i}),
\nonumber \\
P_{b.ka}^{.c} &=&\partial _{a}L_{.bk}^{c}+C_{.bd}^{c}T_{.ka}^{d}-(\delta
_{k}C_{.ba}^{c}+L_{.dk}^{c\
}C_{.ba}^{d}-L_{.bk}^{d}C_{.da}^{c}-L_{.ak}^{d}C_{.bd}^{c}),  \nonumber \\
S_{j.bc}^{.i} &=&\partial _{c}C_{.jb}^{i}-\partial
_{b}C_{.jc}^{i}+C_{.jb}^{h}C_{.hc}^{i}-C_{.jc}^{h}C_{hb}^{i},  \nonumber \\
S_{b.cd}^{.a} &=&\partial _{d}C_{.bc}^{a}-\partial
_{c}C_{.bd}^{a}+C_{.bc}^{e}C_{.ed}^{a}-C_{.bd}^{e}C_{.ec}^{a}  \nonumber
\end{eqnarray}
(the d--curvature components are computed in a similar fashion by using the
formula for curvature coefficients (\ref{curvature})).

\subsection{Einstein equations with holonomic--anholonomic va\-riables}

In this subsection we write and analyze the Einstein equations on
5D (pseudo) Riemannian spacetimes provided with anholonomic frame
structures and associated N--connections.

\subsubsection{Einstein equations with matter sources}

The Ricci tensor $R_{\beta \gamma }=R_{\beta ~\gamma \alpha }^{~\alpha }$
has the d--components
\begin{eqnarray}
R_{ij} &=&R_{i.jk}^{.k},\quad R_{ia}=-^2P_{ia}=-P_{i.ka}^{.k},
\label{dricci} \\
R_{ai} &=&^1P_{ai}=P_{a.ib}^{.b},\quad R_{ab}=S_{a.bc}^{.c}.  \nonumber
\end{eqnarray}
In general, since $^1P_{ai}\neq ~^2P_{ia}$, the Ricci d-tensor is
non-symmetric (this could be with respect to anholonomic frames of
reference). The scalar curvature of the metric d--connection, $%
\overleftarrow{R}=g^{\beta \gamma }R_{\beta \gamma },$ is computed
\begin{equation}
{\overleftarrow{R}}=G^{\alpha \beta }R_{\alpha \beta }=\widehat{R}+S,
\label{dscalar}
\end{equation}
where $\widehat{R}=g^{ij}R_{ij}$ and $S=h^{ab}S_{ab}.$

By substituting (\ref{dricci}) and (\ref{dscalar}) into the 5D Einstein
equations
\begin{equation}
R_{\alpha \beta }-\frac 12g_{\alpha \beta }R=\kappa \Upsilon _{\alpha \beta
},  \label{5einstein}
\end{equation}
where $\kappa $ and $\Upsilon _{\alpha \beta }$ are respectively the
coupling constant and the energy--momentum tensor we obtain the
h-v-decomposition by N--connection of the Einstein equations
\begin{eqnarray}
R_{ij}-\frac 12\left( \widehat{R}+S\right) g_{ij} &=&\kappa \Upsilon _{ij},
\label{einsteq2} \\
S_{ab}-\frac 12\left( \widehat{R}+S\right) h_{ab} &=&\kappa \Upsilon _{ab},
\nonumber \\
^1P_{ai}=\kappa \Upsilon _{ai},\ ^2P_{ia} &=&\kappa \Upsilon _{ia}.
\nonumber
\end{eqnarray}
The definition of matter sources with respect to anholonomic frames is
considered in Refs. \cite{vf,v}.

\subsubsection{5D vacuum Einstein equations}

The vacuum 5D, locally anisotropic gravitational field equations, in
invariant h-- v--components, are written
\begin{eqnarray}
R_{ij} &=&0,S_{ab}=0,  \label{einsteq3} \\
^1P_{ai} &=&0,\ ^2P_{ia}=0.  \nonumber
\end{eqnarray}

The main `trick' of the anholonomic frames method for integrating the
Einstein equations in general relativity and various (super) string and
higher / lower dimension gravitational theories is to find the coefficients $%
N_{j}^{a}$ such that the block matrices $g_{ij}$ and $h_{ab}$ are
diagonalized \cite{vf,v,v1}. This greatly simplifies computations. With
respect to such anholonomic frames the partial derivatives are N--elongated
(locally anisotropic).

\section{Proof of the Theorem 3}

We prove step by step the items of the Theorem 3.

The first statement with respect to the solution of
(\ref{ricci1a}) is a connected with the well known result from 2D
(pseudo)\ Riemannian gravity that every 2D metric can be
redefined by using coordinate transforms into a conformally flat
one.

The equation (\ref{ricci2a}) can be treated as a second order differential
equation on variable $v,$ with parameters $x^i,$ for the unknown function $%
h_5(x^i,v)$ if the value of $h_4(x^i,v)$ is given (or inversely
as a first order differential equation on variable $v,$ with
parameters $x^i,$ for the unknown function $h_4(x^i,v)$ if the
value of $h_5(x^i,v)$ is given). The formulas (\ref{p1}) and
(\ref{p2}) are consequences of integration on $v$ of the equation
(\ref{ricci2a}) being considered also the degenerated cases when
$h_5^{*}=0$ or $h_4^{*}=0.$

Having defined the values $h_{4}$ and $h_{5},$ we can compute the values the
coefficients $\alpha _{i},\beta $ and $\gamma $ (\ref{abc}) and find the
coefficients $w_{i}$ and $n_{i}$ The first set (\ref{w}) for $w_{i}$ is a
solution of three independent first order algebraic equations (\ref{ricci3a}%
) with known coefficients $\alpha _{i}$ and $\beta $. The second set of
solutions (\ref{n}) for $n_{i}$ is found after two integrations on the
anisotropic variable $v$ of the independent equations (\ref{ricci4a}) with
known $\gamma $ (the variables $x^{i}$ being considered as parameters). In
the formulas (\ref{n}) we distinguish also the degenerated cases when $%
h_{5}^{\ast }=0$ or $h_{4}^{\ast }=0.$

Finally, we note that the formula (\ref{confsol}) is a simple
algebraic consequence from (\ref{confeq}).

The Theorem 3 has been proven.

\section{Proof of Theorem 5}

We emphasize the first two items:

\begin{itemize}
\item  The equation (\ref{ricci1const}) imposes a constraint on coefficients
of a diagonal 2D metric parametrized by coordinates $x^{2}=u$ and $%
x^{3}=\lambda .$ By coordinate transforms $x^{2,3}\rightarrow \widetilde{x}%
^{2,3}\left( u,\lambda \right) ,$ see for instance, \cite{petrov} we can
reduce 2D every metric
\[
ds_{[2]}^{2}=g_{2}(u,\lambda )du^{2}+g_{3}(u,\lambda )d\lambda ^{2}
\]
to a conformally flat one
\[
ds_{[2]}^{2}=\varpi (\widetilde{x}^{2},\widetilde{x}^{3})\left[ d(\widetilde{%
x}^{2})^{2}+\epsilon d(\widetilde{x}^{3})^{2}\right] ,\epsilon =\pm 1.
\]
with conformal factor $\varpi
(\widetilde{x}^{2},\widetilde{x}^{3}),$ for which
(\ref{ricci1const}) transforms into (\ref{auxr1}) with new 'dot'
and
'prime' derivatives $\varpi ^{\bullet }=\partial \varpi /\partial \widetilde{%
x}^{2}$ and $\varpi ^{^{\prime }}=\partial \varpi /\partial \widetilde{x}%
^{3}.$ It is not possible to find an explicit form of the general solution
of (\ref{auxr1}). If we approximate, for instance, that $\varpi =\varpi
\left( \widetilde{x}^{2}\right) ,$ the equation
\[
\varpi \varpi ^{\bullet \bullet }-(\varpi ^{\bullet })^{2}=2\Lambda \varpi
^{3}
\]
has an exact solution (see 6.127 in \cite{kamke}) which can be found from a
Bernulli equation
\[
(\varpi ^{\bullet })^{2}=\varpi ^{3}\left( C\varpi ^{-1}+4\Lambda \right)
,C=const,
\]
which allow us to find $\widetilde{x}^{2}(\varpi ),$ or, in non explicit
form $\varpi =\varpi \left( \widetilde{x}^{2}\right) .$ We can chose a such
solution as a background one and by using conformal factors $\Omega (%
\widetilde{x}^{2},\widetilde{x}^{3}),$ transforming $\varpi (\widetilde{x}%
^{2},\widetilde{x}^{3})$ into $\varpi \left( \widetilde{x}^{2}\right) $ we
can generate solutions of the 5D Einstein equations with anisotropic
cosmological constant by inducing second order anisotropy $\zeta _{i}.$ The
case when $\varpi =\varpi \left( \widetilde{x}^{3}\right) $ is to be
obtained in a similar manner by changing the 'dot' derivative into 'prime'
derivative.

\item  The equation (\ref{ricci2const}) das not admit $h_{5}^{\ast }=0$
because in this case we must have $h_{5}=0.$ For a given value of
$h_{5},$
introducing a new variable $\tau =h_{5}^{\ast }/2h_{5}$ we can transform (%
\ref{ricci2const}) into a first order nonlinear equation for $h_{4}$ (\ref
{auxr2p}),which can be transformed \cite{kamke} to a Ricatti, then to a
Bernulli equation which admits exact solutions. We note that the holonomic
coordinates are considered as parameters. The inverse problem, to find $%
h_{5} $ for a given $h_{4}$ is more complex because is connected
with solution of a second order nonlinear differential equation
\begin{equation}
h_{5}^{\ast \ast }+\frac{(h_{5}^{\ast })^{2}}{2h_{5}}-\frac{h_{4}^{\ast }}{%
2h_{4}}h_{5}^{\ast }-2\Lambda h_{4}h_{5}=0,  \label{auxr2p}
\end{equation}
which can not integrated in general form. Nevertheless, a very
general class of solutions can be found explicitly if
$h_{4}^{\ast }=0,$ i. e. if $h_{4}$
depend only on holonomic coordinates. In this case the equation (\ref{auxr2p}%
) can be reduced to a Bernulli equation \cite{kamke} which admits exact
solutions.

\item  The formulas (\ref{aw}), (\ref{nlambda}) and (\ref{aconf4}) solving
respectively (\ref{ricci3const}), (\ref{ricci4const}) and (\ref{confeql})
are proven similarly as for the Theorem 3 with that difference that in the
presence of the cosmological term $h_{5}^{\ast }\neq 0,\beta \neq 0$ and, in
general, $w_{i}\neq 0.$
\end{itemize}

The Theorem 5 has been proven.

\end{document}